\PassOptionsToPackage{prologue,dvipsnames,table}{xcolor}
\documentclass[acmtog]{acmart}

\usepackage{booktabs} 
\usepackage{bbm}
\usepackage{nicefrac}
\usepackage[normalem]{ulem}
\usepackage{empheq} 
\usepackage{algorithm}
\usepackage{algpseudocode}
\usepackage{hyperref}

\usepackage{graphicx}
\usepackage{subcaption}
\usepackage{bm}
\usepackage{xspace}
\usepackage{diagbox}
\usepackage{multirow} 
\usepackage{soul}
\usepackage{changepage,threeparttable} 
\usepackage{enumitem}
\usepackage{minted}
\usepackage{listings}

\AtBeginDocument{%
  }

\setcopyright{acmlicensed}
\acmJournal{TOG}
\acmYear{2025} \acmVolume{44} \acmNumber{6} \acmArticle{216} \acmMonth{12} \acmDOI{10.1145/3763271}

\begin{document}

\title{GarmageNet: A Multimodal Generative Framework for Sewing Pattern Design and Generic Garment Modeling}


\author{Siran Li}
\authornote{Equal contributions.}
\email{lisiran0223@gmail.com}
\affiliation{%
  \institution{Zhejiang Sci-Tech University and Style3D Research}
  \city{Hangzhou}
  \country{China}
}

\author{Ruiyang Liu}
\authornotemark[1]
\authornote{Project lead and corresponding author. \\ This research work is fully conducted at \href{https://www.style3d.com/}{Style3D Research}.}
\email{liuruiyang@linctex.com}
\author{Chen Liu}
\authornotemark[1]
\email{eric.liu@linctex.com}
\affiliation{%
  \institution{State Key Lab of CAD\&CG, Zhejiang University and Style3D Research}
  \city{Hangzhou}
  \country{China}
}

\author{Zhendong Wang}
\email{wang.zhendong.619@gmail.com}
\author{Gaofeng He}
\email{hegaofeng@linctex.com}
\affiliation{%
  \institution{Style3D Research}
  \city{Hangzhou}
  \country{China}
}

\author{Yong-Lu Li}
\email{yonglu_li@sjtu.edu.cn}
\affiliation{%
  \institution{Shanghai Jiao Tong University}
  \city{Shanghai}
  \country{China}
}
\author{Xiaogang Jin}
\email{jin@cad.zju.edu.cn}
\affiliation{%
  \institution{State Key Lab of CAD\&CG, Zhejiang University}
  \city{Hangzhou}
  \country{China}
}
\author{Huamin Wang}
\email{wanghmin@gmail.com}
\affiliation{%
  \institution{Style3D Research}
  \city{Hangzhou}
  \country{China}
}

\begin{abstract}

Realistic digital garment modeling remains a labor-intensive task due to the intricate process of translating 2D sewing patterns into high-fidelity, simulation-ready 3D garments. We introduce \emph{GarmageNet}, a unified generative framework that automates the creation of 2D sewing patterns, the construction of sewing relationships, and the synthesis of 3D garment initializations compatible with physics-based simulation.
Central to our approach is \emph{Garmage}, a novel garment representation that encodes each panel as a structured geometry image, effectively bridging the semantic and geometric gap between 2D structural patterns and 3D garment geometries. Followed by \emph{GarmageNet}, a latent diffusion transformer to synthesize panel-wise geometry images and \emph{GarmageJigsaw}, a neural module for predicting point-to-point sewing connections along panel contours. 
To support training and evaluation, we build \emph{GarmageSet}, a large-scale dataset comprising 14,801 professionally designed garments with detailed structural and style annotations.
Our method demonstrates versatility and efficacy across multiple application scenarios, including scalable garment generation from multi-modal design concepts (text prompts, sketches, photographs), automatic modeling from raw flat sewing patterns, pattern recovery from unstructured point clouds, and progressive garment editing using conventional instructions, laying the foundation for fully automated, production-ready pipelines in digital fashion. Refer to our \href{https://style3d.github.io/garmagenet/}{project page} for open-sourced code and dataset.

\end{abstract}

\ccsdesc[500]{Computing methodologies~Shape modeling}
\ccsdesc[500]{Computing methodologies~Reconstruction}
\ccsdesc[500]{Computing methodologies~Hierarchical representations}
\ccsdesc[500]{Computing methodologies~Shape representations}
\keywords{Garment Modeling, Garment Dataset, Diffusion Generation}

\begin{teaserfigure}
  \includegraphics[width=\textwidth]{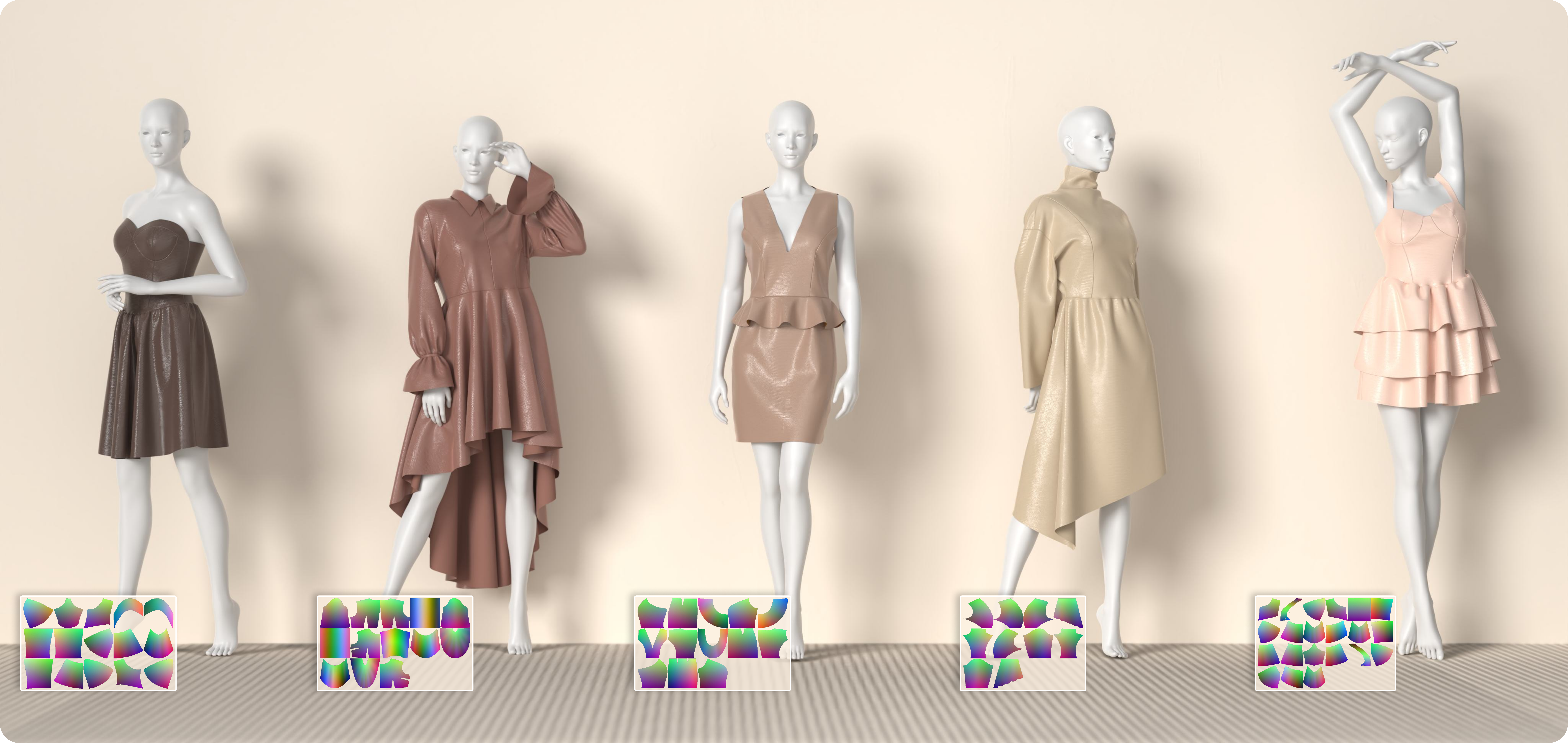}
  \caption{GarmageNet in Action: A diverse and sophisticated collection of garment assets automatically generated by our \emph{GarmageNet} framework, along with their corresponding Garmages—our unified 2D–3D representation that encodes both sewing patterns and detailed geometry for seamless integration with existing garment modeling workflow. Altogether, GarmageNet generates garments across the spectrum of design complexity: from intricate multi-layered ensembles (3rd and 5th) and striking asymmetric styles (2nd and 4th) to form-fitting corsets requiring precise drape and structural fidelity (1st).}
  \label{fig:teaser}
\end{teaserfigure}

\maketitle
\section{Introduction}

Realistic digital clothing plays a pivotal role in entertainment and gaming by enhancing immersion, and in fashion and e-commerce by accelerating product development and reducing costs. Yet, end-to-end 3D garment modeling remains a labor-intensive and technically demanding process, encompassing line-art design, pattern drafting, sewing assignment, 3D initialization~\cite{liu2024automatic}, and physics-based simulation. With the rapid progress of generative AI in 2D fashion design, a pressing question arises: can these models extend beyond image generation to automate the entire fashion manufacturing pipeline, transforming creative designs into production-ready sewing patterns and simulation-ready 3D garments that faithfully mirror their physical counterparts?

Garments, unlike images with well-established 2D representations, are complex 3D forms constructed from two-dimensional sewing patterns. These patterns are not merely cutting templates but encode the rules by which flat fabrics are assembled and folded into complex 3D forms. Effective garment modeling must therefore balance 3D geometric fidelity, to achieve realistic draping, with 2D structural correctness, to ensure manufacturability. Despite recent advances, learning-based garment modeling remains fundamentally limited by trade-offs between structure and geometry.

\begin{figure}[!t]
    \centering
    \includegraphics[width=\linewidth]{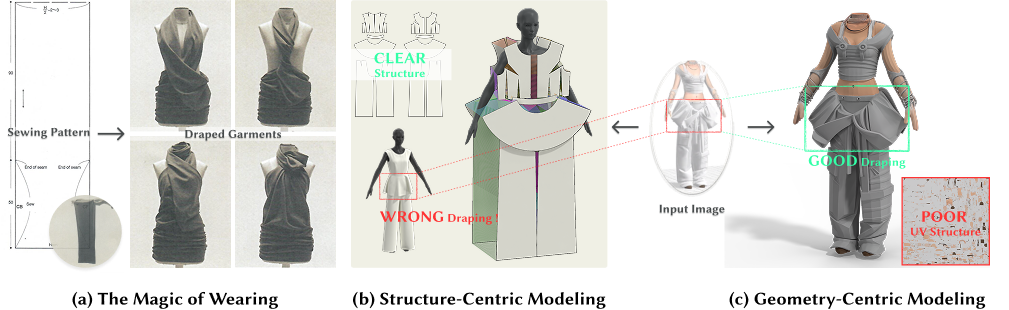}
    \caption{Trade-offs between structure-centric and geometry-centric approaches. The magic of wearing (a) indicates that the same sewing pattern could lead to various draping statuses, raising the problem of \emph{Structure-centric modeling}, which focuses on sewing pattern structure but fails to ensure draping alignment with the input (b). On the other hand, \emph{Geometry-centric modeling} focuses on draping alignment but fails to preserve structure integrity in UV space (c).}
    \label{fig:fwd_vs_bwd}
\end{figure}

\emph{Structure-centric garment modeling methods} operate in the sewing‐pattern domain.
They use auto-regressive or diffusive frameworks to either (i) generate vectorized~\cite{he2024dresscode,nakayama2024aipparel,liu2024sewingldm,li2025garmentdiffusion3dgarmentsewing} or rasterized~\cite{tatsukawa2025garmentimage} sewing patterns together with edge-wise sewing correspondences and 3D initializations as per-panel rigid-transformation matrices, or (ii) emit high-level parameters~\cite{bian2024chatgarment,guo2025garmentx} and programs~\cite{zhou2024design2garmentcode} of a parametric pattern-making DSL such as GarmentCode~\cite{korosteleva2023garmentcode}.
The generated patterns are then draped onto a target avatar using conventional cloth simulation. While this paradigm preserves structural correctness by explicitly producing sewing patterns, it lacks full spatial context and often fails to reproduce fine folds and realistic drape geometry (Figure~\ref{fig:fwd_vs_bwd} (b)).

In contrast, \emph{geometry-centric approaches}~\cite{yu2025surf,rong2024gaussiangarments,zhang2024clay,tochilkin2024triposr,xiang2024structured,zhao2025hunyuan3d} map multi-modal design inputs directly into a draped 3D garment. They typically employ continuous, optimizable implicit representations (e.g., distance or occupancy fields) to maintain geometric fidelity, while failed to incorporate sewing pattern structure which is inherently discrete and discontinuous (Figure~\ref{fig:fwd_vs_bwd} (c)). Not to mention that garment meshes recovered from those contiguous fields often exhibit suboptimal geometric quality, making it extremely challenging, if not impossible, to recover sewing patterns from those 3D models with UV parameterization~\cite{srinivasan2025nuvo,yu2024texgen}. 

These limitations motivate a \emph{foundational garment-modeling framework} that jointly preserves the structural integrity of 2D sewing patterns and the geometric fidelity of 3D drapes.
In response, we present \emph{GarmageNet}, the first unified framework, to our knowledge, that automatically generates 2D sewing patterns, infers sewing relationships, and produces simulation-ready 3D garment initializations from conventional multi-modal inputs. 

At its core is \emph{Garmage} (Figure~\ref{fig:overview} (b)), a novel representation that reformulates a garment asset as a structured collection of per-panel geometry images~\cite{gu2002geometry} in sewing-pattern (UV) space: the alpha channel encodes each panel’s 2D contour, and the RGB channels capture the draped 3D geometry. This design retains the flexibility of standard image formats for seamless integration with image-based algorithms and enables direct reconstruction of simulation-ready 3D garments.

Building on Garmage, \emph{GarmageNet} formulates garment synthesis as a latent diffusion problem. It first learns a compact manifold of admissible panel variations by encoding each cloth piece’s quasi-static 3D geometry and corresponding 2D pattern into a fixed-length latent token. Using this latent space as a strong prior, a diffusion transformer (DiT) learns to assemble tokens into structurally valid garments, yielding a complete Garmage that preserves panel-wise structure while recovering fine-grained cloth-piece geometry.

We then introduce \emph{GarmageJigsaw}, which recovers sewing relations to convert panel-wise Garmages into a simulator-ready seam graph. Unlike edge-pair matching, GarmageJigsaw models sewing relationships as point-to-point correspondences sampled from boundary pixels in the Garmage’s alpha channel, thereby capturing fine-scale folds and pleats. Those correspondences are then consolidated into edge-to-edge constraints via a specially designed heuristic algorithm, ensuring full compatibility with pattern-design and cloth-simulation software.

To further advance GarmageNet toward real-production quality, we present \emph{GarmageSet}, a large-scale dataset comprising 14{,}801 professionally designed, production-ready\footnote{We define “production-ready” as assets derived from real-world manufacturing data (as opposed to purely synthetic datasets such as GarmentCode), conforming to industry standards and fully compatible with existing garment-production workflows.} garments, each represented as a Garmage with fine-grained structural and style annotations.

In summary, our contributions include:

\begin{itemize}
\item We propose \emph{GarmageNet}, the first unified system that \emph{simultaneously} produces structurally correct sewing patterns and vertex-level-precise, simulation-ready 3D garments from various input modalities, including textural descriptions, line-art sketches, commercial photos, unstructured point clouds or flat patterns.

\item We introduce \emph{Garmage}, a novel and compact representation that seamlessly encodes a garment's discrete sewing pattern structure and continuous draping geometry into fixed-length latent tokens, facilitating efficient integration with diffusion-based generative models and multi-modal cross-attention conditioning.

\item We develop \emph{GarmageJigsaw} to recover sewing relationships between Garmage panels by predicting point-to-point stitches along panel contours and convert Garmage into production-ready garment assets, facilitating downstream editing of sewing patterns, material properties, and dynamic simulations.

\item We construct \emph{GarmageSet}, a large-scale dataset of 14{,}801 professionally-crafted garment assets with industrial-grade sewing patterns originated from real manufacturing workflows, alongside detailed structural and style annotations, advancing learning-based garment modeling toward practical usages.

\item Cross-dataset evaluations demonstrate that our framework outperforms both structure-centric and geometry-centric baselines and enables practical applications, including scalable multimodal garment generation, sewing-pattern recovery from point clouds, automatic 3D modeling from flat patterns, and interactive editing via intuitive commands.

\end{itemize}

\section{Related Work}
In traditional physics-based garment modeling, the simulation \cite{wu2022mas,wang2018parallel} accelerated by specialized hardware \cite{he2025ats} is only the final step of a complex pipeline. Prior to simulation, practitioners must design 2D patterns, define sewing relationships, select fabric properties \cite{wang2023sdb,feng2022lbs} and mesh resolutions \cite{zhang2025pecmr}, and carefully initialize the 3D configuration to avoid cloth–body interpenetration and self-collisions \cite{wang2015tightccd,tang2014bsc,wang2016ers,wang2018ascd}, and non-physical draping in garments with intricate structures \cite{guo2025pmkt,guo2025poa}. Early automation addressed isolated substeps. For example, learning-based sewing identification still required manual initialization and bespoke parsers~\cite{berthouzoz2013parsing}, while recent automatic initialization methods presuppose complete, professionally made patterns with correct seam graphs~\cite{liu2024automatic}. These constraints have motivated learning-based approaches, which in turn depend on large, high-quality datasets. Below, we review available garment datasets and the two primary modeling paradigms they enable.

\subsection{Garment Datasets}
\label{sec:related_garmet_dataset}
Learning-based garment modeling draws on three main data sources: scanning-based, simulation-based, and sewing pattern-based.

\subsubsection{Scanning-based datasets}
\label{sec:related_garmet_dataset_scan}
They capture the geometry and appearance of clothed humans using multi-view RGB rigs, RGB-D sensors, or structured light \cite{pons2017clothcap,xiang2020monoclothcap,zhang2017detailed,lin2023leveraging}. Pipelines reconstruct per-frame surfaces via multi-view stereo or volumetric fusion \cite{pons2017clothcap,zhang2017detailed}, then non-rigidly register a common topology (e.g., SMPL~\cite{loper2015smpl} or garment/body templates~\cite{xiu2023econ}) to obtain temporally consistent meshes and UV textures \cite{bhatnagar2019multi,tiwari2020sizer,ma2020learning,wang20244d,lin2023leveraging}. While these datasets excel at capturing realistic drape, wrinkles, and dynamics, isolating semantically meaningful parts from the raw scans remains a labor-intensive process, heavily reliant on manual efforts. As a result, these datasets typically lack sewing patterns that match the garment assets. Additionally, they are mostly derived from existing commercial garment asset libraries, limiting their scale and design diversity.

\subsubsection{Simulation-based datasets}
\label{sec:related_garmet_dataset_sim}
Unlike scanning-based datasets that focus on garment drape but typically lack explicit sewing structure, simulation based datasets start from manually authored garment assets and use physics engines to synthesize draping across diverse body shapes and motion sequences ~\cite{bertiche2020cloth3d, black2023bedlam, gundogdu2019garnet, jiang2020bcnet, patel2020tailornet, zou2023cloth4d, santesteban2019learning,narain2012adaptive}. Because authoring high-quality assets is labor-intensive and these datasets primarily target clothed-human reconstruction, they emphasize automated expansion of pose and appearance variation, while offering limited structural diversity, often only one or two hundreds of distinct garment designs. ~\cite{luo2024garverselod} are among the first to scale structural variety to the thousands (with 2,869 unique designs); however, the assets remain less complex than production garments without considering multilayer constructions, complex seam topologies, and sewing patterns are not provided.

\subsubsection{Sewing pattern datasets}
\label{sec:related_garmet_dataset_sp}
 Compared with scanning or simulation based dataset, sewing pattern datasets remain relatively nascent. GarmentCodeData~\cite{korosteleva2021generating,korosteleva2024garmentcodedata} is the most widely used sewing-pattern dataset in the research community, generated by programmatically sampling parametric sewing-pattern programs (GarmentCode), and draping onto SMPL~\cite{loper2015smpl} avatars via a fast XPBD simulator~\cite{warp2022}. Although its procedural generation enables easy scaling, it substantially simplifies realities of production data. For example, production patterns contain multi-edge-loop panels and intricate contours (e.g., spirals), beyond the single-edge-loop and four edge-type assumptions in GarmentCodeData; In terms of sewing representation, industrial garments require many-to-many, partial-edge mappings to support multi-layering and dense pleats, whereas GarmentCodeData encodes one-to-one full-edge correspondences; and GarmentCodeData ships with per-panel rigid placements that allow immediate draping, while in practice vendors typically provide only flat DXF patterns without seam graphs or rigid placements, making the recovery of simulator-ready initializations a nontrivial task.

\subsection{Learning-based Garment Modeling}
Depending on the garment representations provided by those datasets, learning-based garment modeling methods can be broadly categorized into two families: geometry-centric and structure-centric.

\subsubsection{Geometric-centric Garment Modeling}
\label{sec:related_garmet_modeling_geo}
Implicit garment modeling methods often rely on unsigned distance fields (UDF)~\cite{yu2025surf,chen2024neural}, manifold distance fields~\cite{liu2024gshell}, and Gaussian splatting~\cite{liu2024clothedreamer, rong2024gaussiangarments} to handle non-watertight garment geometry, and employ diffusion or GAN-based generative models to generate visually pleasant garment assets, with vivid dynamics~\cite{xie2024physgaussian,rong2024gaussiangarments}. However, how to transform those implicit representations into triangular or quadrilateral meshes relies on a solid iso-surface extraction algorithm, which remains quite a challenging problem.

Another option is to explicitly deform a template garment mesh and register it to a posed avatar via end-to-end optimization~\cite{zhu2022registering,qiu2023rec} or differentiable simulators~\cite{li2024diffavatar,sarafianos2024garment3dgen}. While these methods are able to preserve the template’s topological integrity, the geometric expressiveness and complexity of the deformed mesh are also constrained by the template itself.

\subsubsection{Structure-centric garment modeling}
\label{sec:related_garmet_modeling_sp}
Contrary to geometry-centric modeling, \emph{structure-centric} approaches first predict sewing patterns and then drape them onto a target avatar with a cloth simulator. Depending on the representation of the sewing pattern, existing methods fall into three families: vector-quantization based, template-based (program or parameter), and image-based encodings.

\paragraph{Vector-quantization based methods} The pioneering NeuralTailor~\cite{korosteleva2022neuraltailor} formulates sewing-pattern reconstruction from unstructured point clouds as sequence prediction by vector-quantizing patterns into 1D tokens and decoding them with an LSTM equipped with point-level attention. Subsequent work broadens the input modalities: DressCode~\cite{he2024dresscode} generates sewing patterns from text, while SewFormer~\cite{liu2023sewformer} predicts them from images, with~\cite{li2025dress} integrates differentiable simulator to further optimize the predicted patterns so the simulated drape better matches the input image.
Recent diffusion-based~\cite{liu2023sewformer,li2025garmentdiffusion3dgarmentsewing} and LLM-based~\cite{nakayama2024aipparel} variants further refine the vector-quantized formulation and architectures to scale to larger, more complex GarmentCodeData and to accommodate diverse modalities.

\paragraph{Template-based methods} Due to the probabilistic nature of neural-network generators, vector-quantization based methods do not guarantee seam correctness and can produce geometric defects such as self-intersections. To mitigate this, template-based methods~\cite{bian2024chatgarment,zhou2024design2garmentcode,guo2025garmentx} exploit the programmatic nature of GarmentCode~\cite{korosteleva2023garmentcode} by generating high-level design parameters or DSL programs, which enforces structural validity and improves pattern quality. Nevertheless, both vector-quantization and template-based approaches inherit simplifying assumptions from GarmentCodeData and still require physics simulation to perceive the final on-body drape.

\paragraph{Image-based modeling} Our Garmage formulation is informed by image-based sewing-pattern modeling that preserves the 2D nature of patterns using explicit~\cite{chen2022structure,tatsukawa2025garmentimage} or implicit~\cite{li2024isp} image-like encodings. Although raster encodings are more redundant than 1D tokenizations, this capacity allows us to embed UV-aligned position maps that depict 3D drape~\cite{li2024garment,li2024reconstruction,li2025single} and allowing convenient integration with recent image generative models~\cite{gu2002geometry,yan2024object,yu2024super}. The added redundancy also necessitates a compact encoding to balance computation and fidelity. We address this with a patch-based geometry-image design that compresses per-panel content while preserving sewing structure, which is a key novelty of Garmage.

\begin{figure*}[t]
    \centering
    \includegraphics[width=\textwidth]{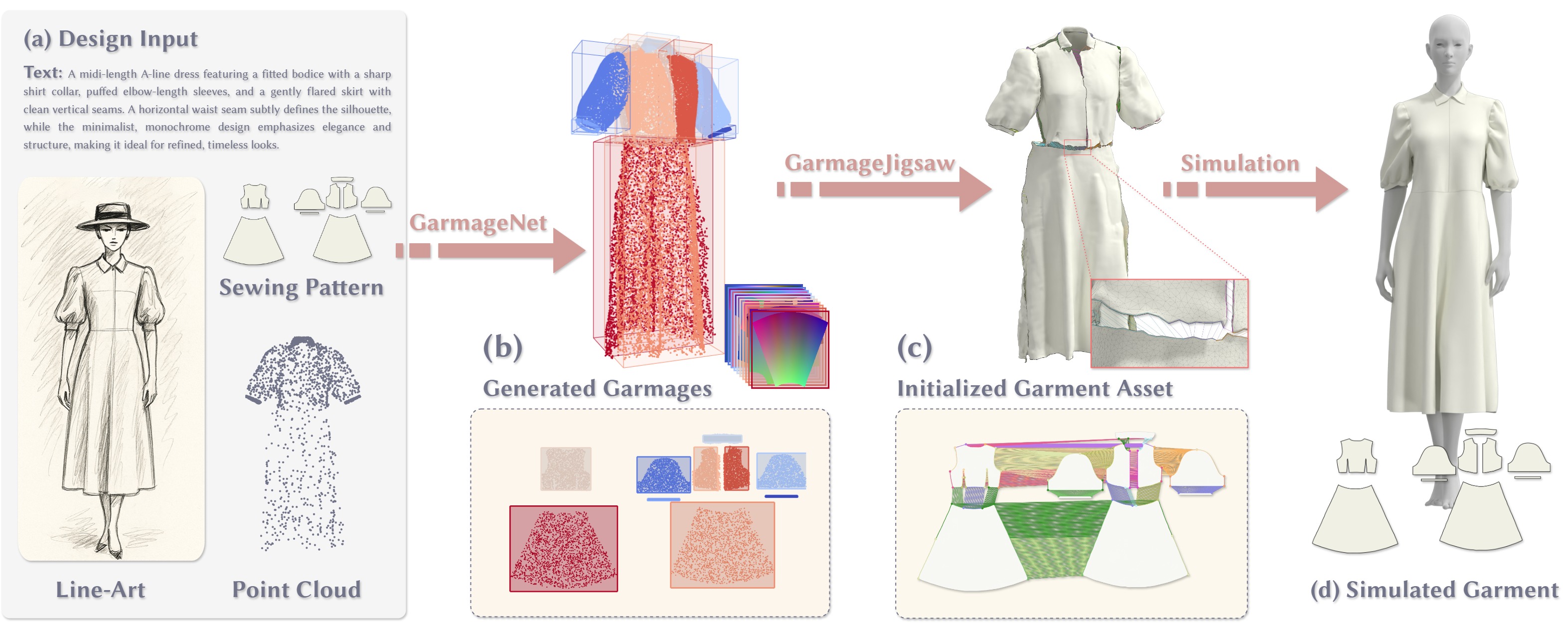}
    \caption{Overview of our \emph{GarmageNet} framework, which seamlessly converts multi-modal design inputs—including text descriptions, sewing patterns, line-art sketches, and point clouds (a)—into simulation-ready garment assets (d). Central to our framework is the novel \emph{Garmage} representation, a 2D–3D unified representation that encodes each garment as a structured set of per-panel geometry images (b). Leveraging \emph{Garmage}, our approach efficiently recovers vertex-level sewing relationships and detailed 3D draping initializations (c), enabling direct and high-quality garment simulation.}
    \Description{}
    \label{fig:overview}
\end{figure*}

\subsection{Structural Object Modeling}
\label{sec:related_structure_gen}
Recent advancements in structural object modeling have enhanced the generation and reconstruction of Boundary Representation (B-rep) models, enabling more complex 3D shape synthesis for CAD applications. BRepGen \cite{xu2024brepgen} uses a diffusion-based approach to generate B-rep models hierarchically, capturing intricate geometries, while SolidGen \cite{jayaraman2022solidgen} employs autoregressive neural networks to predict B-rep components with indexed boundary representation, facilitating high-quality CAD model generation. ComplexGen~\cite{guo2022complexgen} detects geometric primitives and their relationships to create structurally faithful CAD models.

StructureNet \cite{mo2019structurenet}, DPA-Net \cite{yu2025dpa}, and TreeSBA \cite{guo2025treesba} focus on basic geometric shapes but struggle with non-rigid structures and sewing relationships in garments. StructEdit \cite{mo2020structedit} targets local editing of geometric bodies, suitable for regular shapes but limited for flexible, multi-layer garments. 3D Neural Edge Reconstruction \cite{li20243d} reconstructs rigid object contours but does not handle flexible garment modeling.

Fracture assembly methods like PuzzleFusion++ \cite{wang2024puzzlefusion++} and Jigsaw \cite{lu2024jigsaw} infer matching relationships for rigid objects but fail with garments, where misaligned contours and segment-to-segment connections are common. In \emph{Garmage}, we adapt this approach by treating panels as \emph{fractures}, predicting relationships between their contour points to establish sewing connections. Unlike rigid objects, garment contours may not align perfectly, and Garmage addresses this by incorporating garment-specific properties, such as curvature, edge smoothness, and sewing constraints in a learning-based framework.
\section{Overview}

Traditional digital garment modeling demands expert intervention to draft 2D sewing patterns and manually arrange panels around articulated avatars for physics-based simulation. Although learning-based approaches have begun to automate pattern creation, they typically lack explicit 3D geometric guidance, resulting in imprecise outputs and difficulty in handling complex draping behaviors.

We present the first unified learning-based garment creation framework that encodes sewing-pattern structure and its quasi-static drape geometry in a unified image-like representation, \emph{Garmage}.
As shown in Figure~\ref{fig:overview}, the system ingests multimodal design inputs, including text descriptions, sketches, scanned point clouds, and raw sewing patterns, and synthesizes Garmages with a diffusion transformer (DiT).

After generation, we extract 2D sewing contours by thresholding the alpha channel of the generated Garmage and reconstruct each cloth panel’s 3D shape by denormalizing the Garmage's RGB channels with the panel’s bounding box. To recover stitching topology, our GarmageJigsaw module jointly leverages 2D silhouette features and 3D proximity of contour points, producing point correspondences that we fit with B\'ezier curves to yield manufacturable seam lines.
As a result, our method extracts, from the generated Garmage, the garment’s 2D sewing patterns, sewing relationships, and vertex-wise fine-grained initial 3D geometry positioned around a digital avatar—ready for physics-based simulation.

The effectiveness of our framework depends on a comprehensive, multimodal garment corpus, which we call GarmageSet. To support all input pathways, each entry in GarmageSet pairs a ground-truth Garmage (with per-panel geometry images, dimensions, and bounding boxes) with four aligned modalities: a natural-language description, a line-art sketch, a vectorized sewing pattern, and a uniformly sampled point cloud of the draped garment. GarmageSet contains 14{,}801 garments across varied styles and categories, and provides the rich supervision necessary to train and evaluate GarmageNet.
\section{GarmageNet}
\label{sec:garmage}

While intertwining 3D geometry and 2D structure pose challenges for learning-based garment encoding and generation, it confers a unique advantage: garment assets inherently possess well-structured and semantically meaningful UV spaces (i.e., their sewing patterns). Each texel in a sewing-pattern panel simultaneously maps to a point on its corresponding 3D cloth piece, creating a natural bridge between structure and geometry. Garmage exploits this insight by converting each cloth piece into a panel-aligned geometry image, whose color channels encode the piece’s quasi-static 3D geometry and whose alpha channel delineates the panel contour. By harnessing the complementary strengths of 2D and 3D representations, Garmage enables efficient, high-quality 3D garment creation without sacrificing structural fidelity.

\subsection{The Garmage Representation}
\label{sec:garmage-repr}
Traditionally, a garment asset is represented as a set of 3D cloth pieces $\mathcal{C}=\{C_i\}_{i=1}^N$ and their corresponding 2D sewing pattern panels $\mathcal{P}=\{P_i\}_{i=1}^N$, where each panel in the sewing pattern maps directly to a cloth piece in 3D space. 

In Garmage, we model a garment as a set of per-panel geometry images~\cite{gu2002geometry,sander2003multi}, each of which simultaneously encodes a panel’s 2D contour and its normalized 3D shape. Formally, we define:

\begin{equation}\label{eq:garmage_repr}
    \mathcal{G}=  \bigl\{(P_i,C_i)\bigr\}_{i=1}^N = \bigl\{(D_i, B_i, I_i)\bigr\}_{i=1}^N,
\end{equation}
where $D_i=(h_i,w_i)\in\mathbb{R}^2$ gives the $i-$the panel’s physical height and width dimension (in meters) aligned with the fabric’s warp and weft; $B_i=(o_i,s_i)\in\mathbb{R}^{6}$ specifies the axis-aligned bounding box of cloth piece $C_i$, parametrized by its center $o_i\in\mathbb{R}^3$ and half-extents $s_i\in\mathbb{R}^3$;
and $I_i\in\mathbb{R}^{H\times W \times4}$ is a 4-channel image patch whose first three channels encode $C_i$'s geometry normalized by $B_i$ and whose alpha channel delineates the panel contour as an occupancy map.

To construct each image patch $I_i$, we rasterize the cloth piece $C_i$ under its panel $P_i$'s UV parameterization at a uniform resolution $(H=W=256)$. Before rasterization, we rotate the 2D panel $P_i$ so its warp direction aligns with the $v^{+}$ axis, then normalize its coordinates to $[-1, 1]$ by its physical dimension $D_i=(h_i,w_i)$. Simultaneously, we map every 3D vertex $v_j\in C_i$ into normalized space via $(v_j-o_i)/s_i\in[-1, 1]^3$ using its bounding box $B_i=(o_i,s_i)$. At each pixel center $u_p\in[-1, 1]^2$ (corresponding to a 3D point $p$), we test whether $p$ falls inside any triangle of the cloth piece; if so, we find the containing triangle's vertices $j$ and their barycentric weights $\beta_j(p)\in\mathbb{R}^3$ to $p$, then set the pixel value as the sum of the normalized position of those vertices weighted by the barycentric weights:
\begin{equation}
    I_i(u_p) =
    \begin{cases}
        \Bigl(\sum_j\beta_j(u_p)\frac{v_j-o_i}{s_i},\,&1\Bigr), \quad p\in C_i, \\
        \Bigl((0, 0, 0), &0\Bigr),\quad\text{otherwise.} \\
    \end{cases}
    \label{eq:geo_rast}
\end{equation}

Rasterizing panels with sharp features (e.g., dart tips) requires careful handling of boundary aliasing. Following~\cite{yan2024object}, we run the rasterization at an initial high $1024\times 1024$ resolution and subsequently downsample to the target resolution $256\times256$ via sparse pooling.

\begin{figure}[t]
    \centering
    \includegraphics[width=\linewidth]{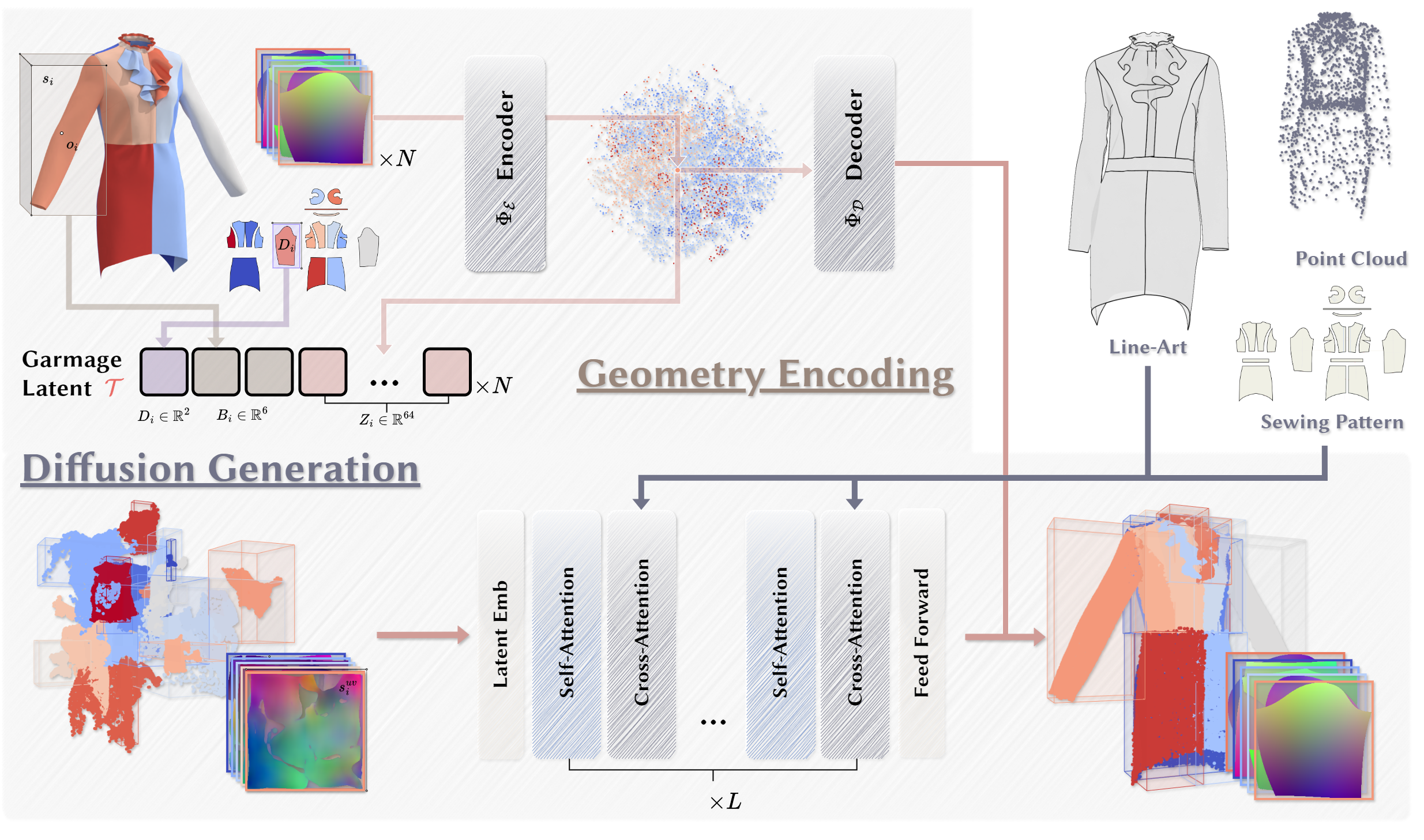}
    \caption{Overview of our \emph{GarmageNet} architecture. During the \emph{geometry encoding} stage (top), each garment is encoded into a set of fixed-size (72-dimensional) latent vectors using a Variational Autoencoder (VAE). These compact latent representations serve as training targets for the subsequent \emph{diffusion generation} stage (bottom). In the diffusion stage, we employ a diffusion transformer (DiT) denoiser, integrating multi-modal conditions, including line-art sketches, raw sewing patterns, and point clouds via cross-attention mechanisms to effectively guide and control the garment generation process.}
    \Description{}
    \label{fig:garmage_pipline}
\end{figure}

\subsection{Diffusion-based Garmage Generation}
\label{sec:garmagenet}
Garment panels serving similar functions often exhibit similar shape features and consistent spatial relationships relative to the human body. For instance, bodice panels typically feature characteristic structural elements such as necklines and armholes and are generally positioned over the chest region in 3D space. This regularity implies a strong correlation between a panel’s silhouette (encoded in the alpha channel of $I_i$) and the geometry status of its corresponding cloth piece (the remaining channels in $I_i$), motivating the compression of Garmages into a unified latent space that simultaneously captures both silhouette and geometry.

\subsubsection{Latent Encoding.}
\label{sec:latent-enc}
Consider the geometry-image component $I_i$ of the $i-$th panel in a Garamge $\mathcal{G}$, we leverage UNet-based variational autoencoder $\bigl(\Phi_\mathcal{E}(\cdot),\Phi_\mathcal{D}(\cdot)\bigr)$, to compress $I_i$ into a 64-dimensional latent vector:
\begin{equation}
\begin{split}
     \Phi_{\mathcal{E}}:&\,\mathbb{R}^{256\times256\times4} \to \mathbb{R}^{64},\quad \Phi_{\mathcal{E}}(I_i)=Z_i, \\
    \Phi_{\mathcal{D}}:&\,\mathbb{R}^{64}\to \mathbb{R}^{256\times256\times4},\quad \Phi_{\mathcal{D}}(Z_i)\approx I_i.        
\end{split}
\end{equation}

To further reinforce 2D–3D correlation in the latent space, during training, we randomly mask out the geometry channels of $I_i$ with probability $0.25$ before passing it through the encoder $\Phi_\mathcal{E}(\cdot)$, and forcing the decoder $\Phi_{\mathcal{D}}(\cdot)$ to reconstruct the geometry part solely from the panel's silhouette during inference. This masking scheme also enables flexible Garmage generation from raw sewing patterns.

After latent compression, any garment can be represented as a set of fixed-length latent tokens:
\begin{equation}
    \mathcal{G}=\mathcal{T} = \bigl\{T_i\bigr\}_{i=1}^N = \bigl\{(D_i\oplus B_i\oplus Z_i)\bigr\}_{i=1}^N\in\mathbb{R}^{N\times 72},
\end{equation}
where $N$ represents the number of panels in each garment, $D_i\in\mathbb{R}^2$ represents the 2D physical dimension of the panel, $B_i\in\mathbb{R}^6$ is the axis-aligned bounding box of its corresponding cloth piece and $Z_i\in\mathbb{R}^{64}$ is the geometry latent.

We train the autoencoder with MSE loss to minimize the reconstruction error, along with a low-weighted ($\lambda_{reg}=1e-6$) KL divergence term between the encoder's approximate posterior $q_{\Phi_{E}}(z|I)$ and a standard normal distribution $p(z)=\mathcal{N}(0,1)$:
\begin{equation}
    \mathcal{L}_{enc} =\frac{1}{N}\sum_{i=1}^N \|I_i-\Phi_{\mathcal{D}}(z_i)\|_2^2 \,+\, \lambda_{reg}D_{KL}\bigl[q_{\Phi_{\mathcal{E}}}(z|I_i)\,\|\,p(z)\bigr].
\end{equation}

\subsubsection{Diffusion Generation}
\label{sec:garmagenet-hiergen}
Based on the learned Garmage latent space, we train a diffusion transformer (DiT) to map random samples from the standard normal distribution $\epsilon\backsim\mathcal{N}(0,1)$ to valid Garmages based on various user input conditions $c$. Specifically, in the forward process, we gradually interpolate the input token $\mathcal{T}$ with random noise through $0\leq t\leq 1000$ timesteps turning it into noisy states $\mathcal{T}_t=\sqrt{\alpha_t}\mathcal{T}_0+\sqrt{1-\alpha_t}\epsilon_t$ at each timestep. In the backward process, we linearly embeds the noised latent $\{Z_{i,t}\}_{i=1}^N$ into patch tokens, embeds the 2D dimension and 3D bounding box $\{D_{i,t}\oplus B_{i,t}\}_{i=1}^N$ into position tokens, and add them together with the embedded timesteps to construct the noisy state, and train the diffusion transformer $\Psi_\mathcal{G}(\cdot)$ to recover the added noise from the previous timestep $t-1$ to $t$, conditioned on the input condition $c$:
\begin{equation}\label{eq:diffusion}
\begin{split}
    & \Psi_{\mathcal{G}}:\mathbb{R}^{N\times 72}\to \mathbb{R}^{N\times 72},\quad \Psi_{\mathcal{G}}(\mathcal{T}_t,t,c)\approx\epsilon_{t}, \\
    & \Psi_{\mathcal{G}}(\mathcal{T}_t,t,c) = \textbf{DiT}\Bigl(\textbf{PosEmb}(\mathbf{D}_t\oplus\mathbf{B}_t)+\textbf{MLP}(\mathbf{Z}_t), \,\, t, c\Bigr), \\
    &\mathbf{D}_t\oplus\mathbf{B}_t=\{D_{i,t}\oplus B_{i,t}\}_{i=1}^N\in\mathbb{R}^{N\times 8},\;
    \mathbf{Z}_t=\{Z_{i,t}\}_{i=1}^N\in\mathbb{R}^{N\times 64}.
\end{split}
\end{equation}
We train $\Psi_{\mathcal{G}}(\cdot)$ using mean-squared error between the predicted noise $\Psi_{\mathcal{G}}(\mathcal{T}_0,c,t)$ and added noise $\epsilon_t$: 
\begin{equation}
    \mathcal{L}_{\Psi} = \mathbb{E}_{t,\mathcal{T}_0,\epsilon_t}\Bigl[ \|\epsilon - \Psi_\mathcal{G}(\mathcal{T}_0,c,t)\|_2^2 \Bigr],
\end{equation}
where $\epsilon_t$ denotes the Gaussian noise added at timestep $t$. 

All panels in a Garmage are denoised in parallel while the self-attention mechanism of the Transformer backbone implicitly models the connections between panels, ensuring structural validity of the generated garment. For convenience, we zero‑pad each garment to have a fixed number of panels $N=32$ during training and discarding any panels whose bounding box volume $|B_i|<0.075$ or 2D dimension $\|D_i\|^2<1e^{-4}$ at inference time to accommodate panel number variance.

\subsubsection{Dealing with Conditions}
\label{sec:cond_gen}

During diffusion training, each design modality is first encoded into its own latent space by a pretrained encoder and then injected into the Garmage denoiser via cross-attention. 
\emph{Text prompts} are mapped to a 1024-dimensional text latent using the CLIP text encoder; 
\emph{Line-art sketches} are passed through a pretrained DINOv2 vision transformer, also yielding a 1024-dimensional image latent;
and \emph{unstructured point clouds} are processed by a PointTransformer v3 (PTv3) fine-tuned on GarmageSet for panel segmentation tasks, producing a 1024-dimensional point latent~\cite{wu2024ptv3}.

Unlike other modalities, raw sewing pattern conditioned generation is natively supported by GarmageNet via our VAE’s masking scheme (Section~\ref{sec:latent-enc}), which allows for inferring the full 4-channel geometry images from the silhouettes alone.
Consequently, when a \emph{sewing pattern} is provided, its 2D dimensions $\mathbf{D}_0$ and geometry latents $\mathbf{Z}_0$ are known a prior, leaving only the 3D bounding‐boxes $\mathbf{B}_0$ to be recovered via diffusion. In practice, at each diffusion step $t$, we corrupt $\mathbf{D}_0$ and $\mathbf{Z}_0$ to their noised version $\mathbf{D}_t$, $\mathbf{Z}_t$ at timestep $t$, concatenate them with $\mathbf{B}_t$, and allow the network to iteratively denoise $\mathbf{B}_t$ toward the desired 3D bounding-boxes $\mathbf{B}_0$.

It is worth noting that although Garmage builds on geometry images, our panel-wise geometry-image design enables more efficient latent compression. Combined with the GarmageNet architecture, which explicitly models inter-panel spatial and connectivity relations, the framework achieves substantial gains in generation quality and efficiency compared with the multi-chart geometry-image baseline Omage~\cite{yan2024object} (Table~\ref{tab:quality_comp}).
\begin{figure}[t]
    \centering
    \includegraphics[width=\linewidth]{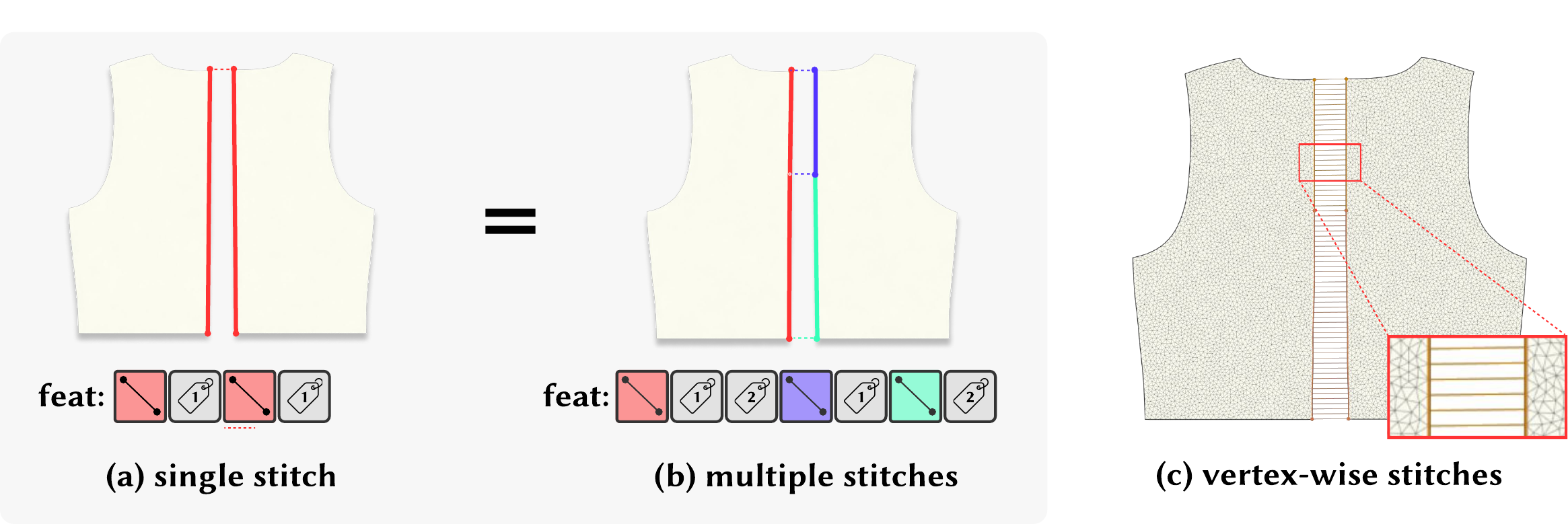}
    \caption{Illustration of stitch representation ambiguity and our point‐wise solution. Existing edge‐based methods suffer from inconsistencies due to arbitrary edge splits: in (a) and (b), the red lines depict the same physical stitch, yet their extracted edge features (shown below) differ in both length and parameter encoding. In contrast, our point‐wise stitching (c) directly anchors stitch correspondences to mesh vertices in physical space, producing consistent, robust sewing relationships independent of panel tessellation.}
    \label{fig:sewing_problem}
\end{figure}

\section{Garamge Processing}

With GarmageNet, we can synthesize complete Garmages from conventional design input, the next challenge is to integrate these rasterized panel images into traditional garment‐modeling pipelines, which requires vectorizing panel contours and reestablishing sewing relationships. 
In the following section, we introduce GarmageJigsaw, a dedicated module that leverages Garmage’s embedded 2D silhouettes and 3D spatial cues to robustly infer vertex-wise sewing correspondences, followed by post‐processing routines that yield production‐ready, vector‐format sewing patterns.

\subsection{Boundary Point Sampling} 
\label{sec:garmage-process}

Conventional vector-quantization based garment modeling systems (Sec. ~\ref{sec:related_garmet_modeling_sp}) define sewing relationships as continuous curve-to-curve correspondences. While straightforward, this edge-based scheme often introduces ambiguity due to ill-defined edge separation and complex many-to-many mappings (Figure~\ref{fig:sewing_problem}), which can cause vector-quantization based methods to fail to converge (see Exp.2 in Table~\ref{tab:cross_dataset_comp}).
Therefore, we instead represent sewing as connectivity between boundary vertices of cloth pieces, an atomic stitch representation that is not subject to further subdivision.

As discussed before, in Garmage, each panel is represented by a four-channel image patch $I_i\in\mathbb{R}^{256\times 256\times 4}$, whose alpha channel \([I_i]_4\) delineates the panel contour. We extract the set of 2D contour points as:
\begin{equation}\label{eq:boundary_extraction}
\begin{split}
    & \partial I_i \in \mathbb{R}^{k_i\times2}, \text{ and }\\
    & \partial I_i=\bigl\{\,
    u_p : [I_i]_4(u_p)>0\bigr\}
\;\setminus\;
    \bigl\{\,u : \bigl([I_i]_4 \ominus \Lambda\bigr)(u_p)>0\bigr\},
\end{split}
\end{equation}
where \(\ominus \Lambda\) denotes binary erosion with elliptical structuring element $\Lambda$ of size $(3,3)$, and $k_i$ refers to the number of contour points from image patch $I_i$.
Denoting \([I_i]_{0:3}\) as the remaining geometric channels from $I_i$, we retrieve the corresponding 3D points for \(\partial I_i\) from \([I_i]_{0:3}\) and denormalize them into world coordinate with the panel's corresponding bounding box \(B_i\):

\begin{equation}
    \rho I_i\in\mathbb{R}^{k_i\times 3},\,\, \rho I_i=\textbf{Denorm}\bigl(\,[I_i]_{0:3}(\partial I_i),\,\, B_i \, \bigr).
\end{equation}

Note that panels may yield nonuniform point densities, we apply resampling under predefined particle distance to \(\rho I_i\), ensuring that adjacent contour samples across all panels exhibit consistent 3D distances, producing normalized inputs for our \emph{GarmageJigsaw} correspondence module.

\begin{figure}[t]
    \centering
    \includegraphics[width=\linewidth]{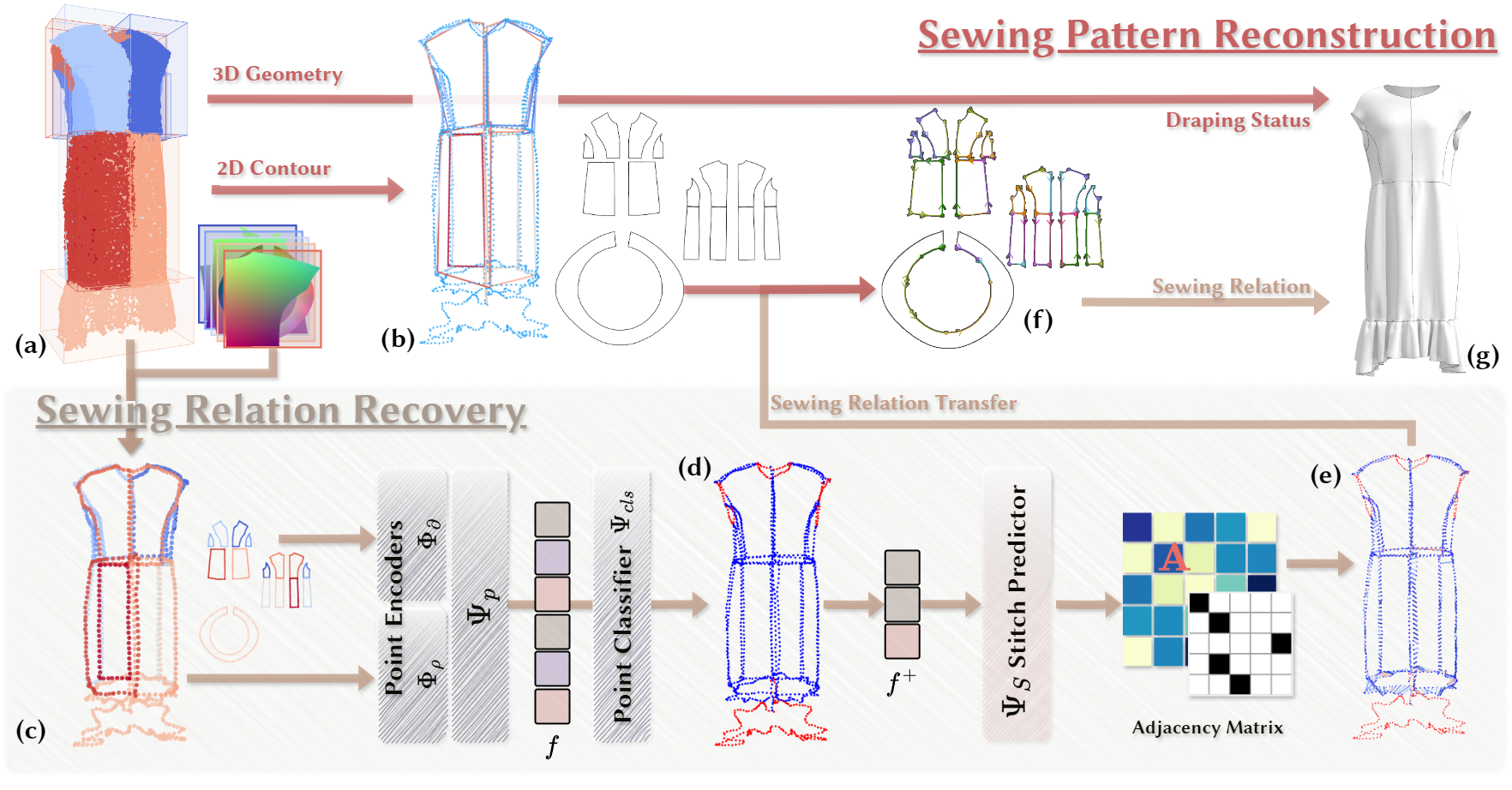}
    \caption{Overview of sewing relationship recovery and simulation-ready sewing pattern reconstruction from the generated \emph{Garmage} (a).  Unlike previous edge-based methods, we predict vertex-level sewing relationships. Specifically, we first sample boundary points (c) from the generated Garmage representation. Our \emph{GarmageJigsaw} takes the boundary points as input, and leverages a point classifier to identify sewing versus non-sewing points (d), followed by a stitch predictor that recovers point-to-point stitches (e), represented as an adjacency matrix. Concurrently, we extract vectorized sewing patterns (b) from the Garmage and transfer the predicted point stitches onto these vectorized patterns (f). We then reconstruct triangle meshes from the vectorized sewing pattern with a Delaunay triangulation constraint by the predicted stitches. Finally, we retrieve vertex-wise draping status from the generated Garmage, leading to a simulation-ready triangle mesh that can be directly integrated into any conventional cloth simulation engine to produce the physically plausible garment (g).}
    \Description{}
    \label{fig:garmagejigsaw_pipline}
\end{figure}

\subsection{Sewing Relation Recovery}
\label{sec:garmagejigsaw}

With the resampled contour points, \emph{GarmageJigsaw} recovers point‐to‐point sewing by jointly leveraging 2D silhouette and 3D geometric features. As shown in Figure ~\ref{fig:garmagejigsaw_pipline}, we first extract per‐point features using two PointNet++ encoders, \(\Phi_{\rho}(\cdot)\) on the 3D contour points \(\rho \mathbf{I}\in\mathbb{R}^{K\times3}\) and \(\Phi_{\partial}(\cdot)\) on the 2D pixels \(\partial \mathbf{I}\in\mathbb{R}^{K\times2}\). These features are concatenated and fused through a series of point‐transformer blocks \(\Psi_{p}(\cdot)\) to yield a \(128\)-dimensional per-point feature matrix:

\begin{equation}
    \begin{split}
        & f\in \mathbb{R}^{K\times 128},\quad f = \Psi_p\Bigl(\,\Phi_{\rho}(\rho \mathbf{I})\,\oplus\,\Phi_{\partial}(\partial \mathbf{I})\,\Bigr), \\
        & \rho\mathbf{I}=\{\rho I_i\}_{i=1}^N\in \mathbb{R}^{K\times 3} \;\text{and}\;
        \partial\mathbf{I}=\{\partial I_i\}_{i=1}^N\in\mathbb{R}^{K \times 2}.
    \end{split}
\end{equation}
Here, \(K=\sum_i k_i\) is the total number of contour points across all panels. A point classifier head \(\Phi_{\mathrm{cls}}(\cdot)\) then selects the subset \(f^+\in\mathbb{R}^{K^+\times128}\) of candidate sewing points by predicting sewing probability based on the point features $f$.

\begin{figure}[t]
    \centering
    \includegraphics[width=\linewidth]{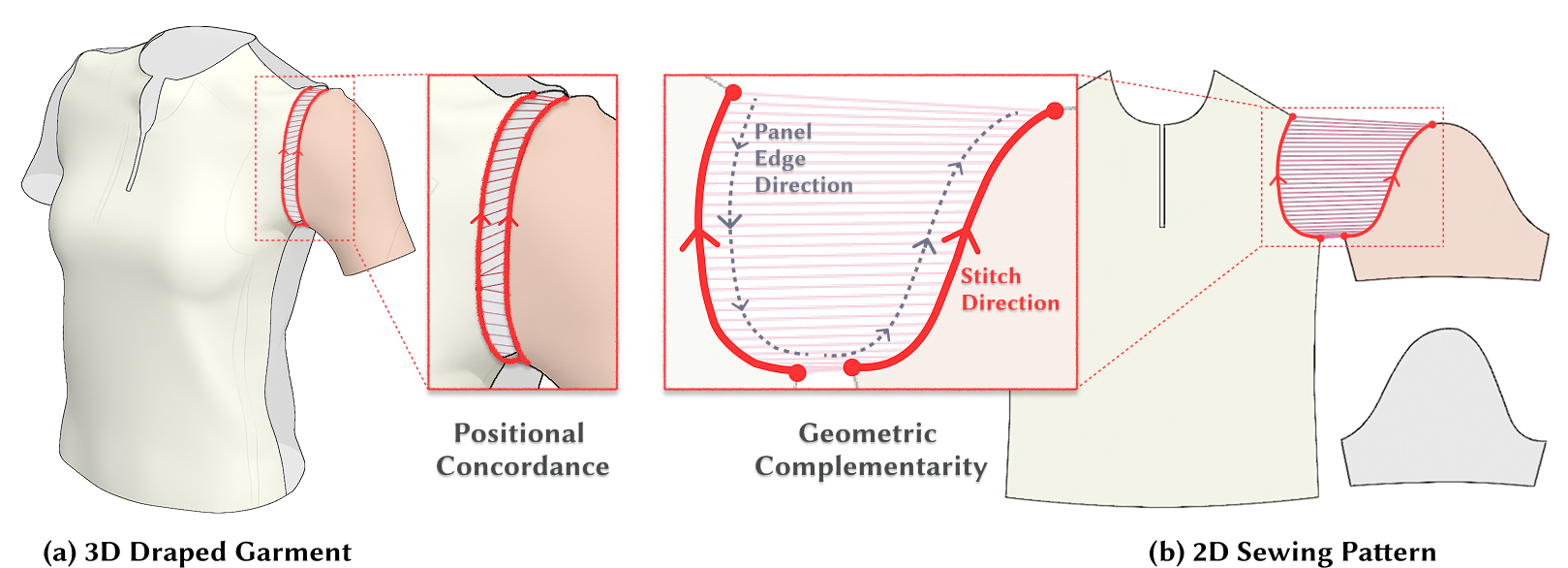}
    \caption{Positional concordance and geometric complementarity of stitched edge pairs. At the armhole, the bodice armscye curve ($\mathbf{AS}$) and the sleeve cap curve ($\mathbf{SC}$) lie in nearly the same 3D location on the draped garment (a), but in 2D sewing pattern space they are complementary in shape: The armscye curve $\hat{\mathbf{AS}}$ is predominantly concave while the sleeve cap curve ($\hat{\mathbf{SC}}$) is convex. Moreover, their stitch orientation opposes each panel’s edge-winding direction: $\hat{\mathbf{SC}}$ is traced with the sleeve panel edge direction, whereas $\hat{\mathbf{AS}}$ is traced against the bodice panel edge direction.}
    \label{fig:primal_dual}
\end{figure}

Seam correspondence between two panels is governed by both 3D positional concordance and 2D geometric complementarity (Figure~\ref{fig:primal_dual}). To capture both aspects, we follow the primal/dual disentanglement in Jigsaw~\cite{lu2024jigsaw}, and attach two lightweight MLP heads, $\Phi_{\mathrm{prime}}(\cdot)$ and $\Phi_{\mathrm{dual}}(\cdot)$, to produce for each boundary point a pair of embeddings that disentangle the from-panel view from its complementary counterpart,thereby suppressing self-matches and favoring complementary correspondences:

\begin{equation}
    \begin{split}
    & f^+_{\mathrm{prime}}\in\mathbb{R}^{K^+\times 128},\,\,f^+_{\mathrm{prime}} = \Phi_{\mathrm{prime}}(f^+), \\
    & f^+_{\mathrm{dual}}\in\mathbb{R}^{K^+\times 128},\,\,f^+_{\mathrm{dual}} = \Phi_{\mathrm{dual}}(f^+),
    \end{split}
\end{equation}

We then combine the primal/dual feature with a learnable symmetric weight matrix $\Lambda_{\mathbf{A}}\in\mathbb{R}^{128\times128}$, followed by a Sinkhorn normalization \cite{cuturi2013sinkhorn} to produce the adjacency probability matrix:
\begin{equation}
    \mathbf{A} \;=\;\textbf{Sinkhorn}\biggl(\exp \Bigl(\frac{(f^+_{\mathrm{prime}})^\top\;\Lambda_{\mathbf{A}}\;f^+_{\mathrm{dual}}}{\tau}\Bigr)\biggr)\;\in\;[0,1]^{K^+\times K^+}.
\end{equation}
where \(A_{ij}\approx A_{j,i}\) denotes the probability of a sewing exists between the \(i\)-th and \(j\)-th contour points, and $\tau$ is a temperature parameter according to ~\cite{lu2024jigsaw}. The probability matrix $\mathbf{A}$ is processed with the Hungarian algorithm~\cite{fischler1981random}, yielding the final point‐to‐point correspondences for seam reconstruction.

The entire GarmageJigsaw model is trained end-to-end with two complementary loss terms: a binary cross-entropy loss \(\mathcal{L}_{\mathrm{cls}}\) that supervises the predicted sewing-point probabilities against ground-truth labels \(y_i\in\{0,1\}\), and a matching loss \(\mathcal{L}_{\mathrm{match}}\) that aligns the predicted adjacency matrix \(\mathbf{A}\) with the ground-truth matrix \(\mathbf{A}_{gt}\). Notably, to prevent the network from trivially minimizing \(\mathcal{L}_{\mathrm{match}}\) by omitting sewing pairs, we pad the ground-truth adjacency matrix \(\mathbf{A}_{gt}\) into a \(K\times K\) with zero columns and rows corresponding to non-sewing points, and compute the matching loss on the whole contour points set \(\bigl(\rho\mathbf{I},\partial\mathbf{I}\bigr)\).

We train GarmageJigsaw with vertex-wise seam supervision, where each stitch is a tuple of vertex indices $(v_a, v_b)$ (Sec.\ref{sec:data_formation}). In the ground-truth assets, stitched vertex pairs are exactly coincident, yielding zero 3D Euclidean distance. By contrast, GarmageNet generated Garmages can exhibit small seam gaps (Figure\ref{fig:overview} (c)) due to residual noise in panel position and scale left by diffusion denoising, as well as the erosion used during boundary extraction (Eq.~\ref{eq:boundary_extraction}). To improve robustness to such gaps, we apply the following augmentations to the ground-truth sewn vertex pairs during training: Firstly, we apply slight translations and scalings to each panel in both 3D space and 2D sewing-pattern space; then inwardly offset boundary facets by a random amount between 2 and 8 mm; and finally add anisotropic noise aligned with seam directions at stitched boundary vertices and isotropic noise at non-stitched boundary vertices.

\begin{figure*}[t]
    \centering
    \includegraphics[width=\linewidth]{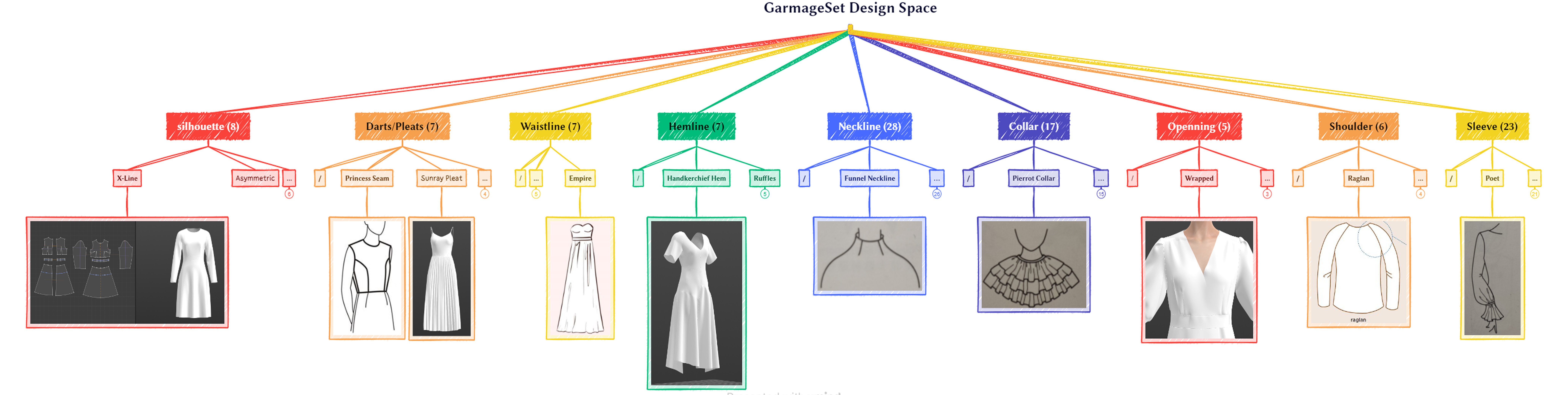}
    \caption{Design space for GarmentSet. Each garment in GarmageSet is annotated along nine professionally defined design dimensions, including \emph{silhouette} (8 options), \emph{darts/pleats} (7), \emph{waistline} (7), \emph{hemline} (7), \emph{neckline} (28), \emph{collar} (17), \emph{opening} (5), \emph{shoulder} (6), and \emph{sleeve} (23). Except for silhouettes, most of those design dimensions have a “/” option indicating that a particular dimension does not apply to the given garment (e.g., sleeve types are irrelevant for skirts or pants).}
    \label{fig:data_design_space}
\end{figure*}

\begin{figure*}[t]
    \centering
    \includegraphics[width=\linewidth]{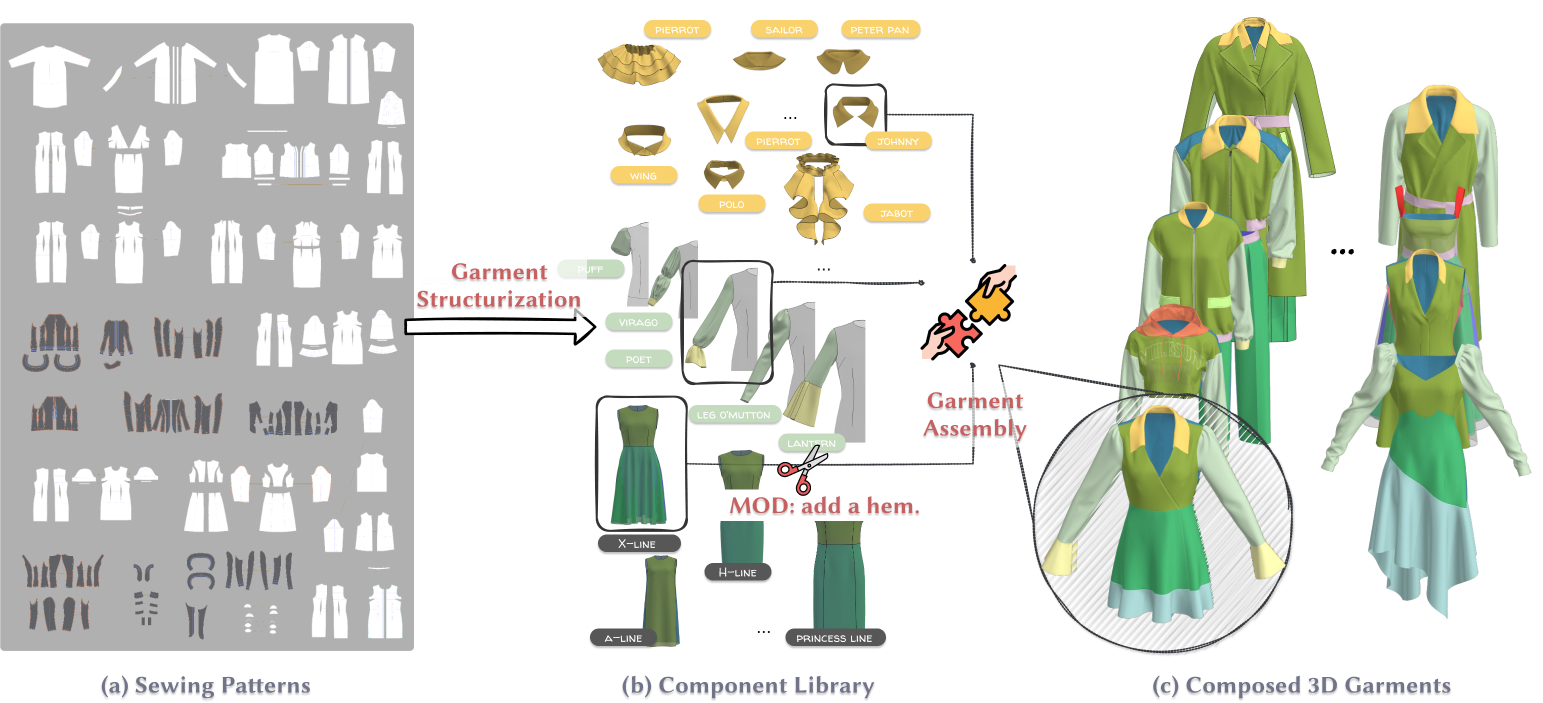}
    \caption{Overview of our \emph{GarmageSet} construction process. We first build a component library (b) by structuring sewing patterns collected in the wild (a). Professional modelers then randomly select several components from this library, apply design modifications such as adjusting width, length, or adding decorative details, and assemble them to create diverse, composed 3D garments (c). This approach enables efficient construction of a high-quality dataset capturing extensive design variability and structural complexity.}
    \label{fig:data_prep}
\end{figure*}

\subsection{Sewing Pattern Reconstruction}
\label{sec:sp-recon}

In conventional garment-modeling workflows, sewing patterns are represented as vectorized curves, with sewing relationships explicitly defined as pairs between these curve segments. To integrate Garmage-generated results seamlessly into existing pipelines, we must vectorize the panel contours and convert the predicted point-to-point sewings into curve-to-curve correspondences.

To vectorize the Garmage panel contours, we first detect corner points exhibiting sharp turning angles along the contour point set $\partial\mathbf{I}$ by employing a specially designed 1D convolutional filter. We then fit piecewise B-spline curves to contour points between adjacent corners, resulting in smooth, compact vector representations for each panel. This vectorization process effectively smooths slanted boundaries (e.g., the last panel of the $4$-th garment in Figure~\ref{fig:teaser}) and fills small noisy holes (e.g., the $2$-nd panel of the $5$-th garment in Figure~\ref{fig:teaser}). Subsequently, we employ a heuristic algorithm to cluster point-to-point stitches predicted by GarmageJigsaw into curve-level sewing correspondences directly on these vectorized B-spline segments.

Finally, we triangulate each sewing-pattern panel into a mesh using constrained Delaunay triangulation~\cite{rognant1999delaunay}, guided by the vectorized panel contours and inferred sewing relationships. Specifically, the boundary facets of each cloth piece mesh consist of contour points uniformly resampled according to the sewing correspondences, ensuring smooth and well-aligned seams between adjacent panels. 

Vertex positions for these triangulated meshes are determined by sampling the corresponding 3D coordinates from their associated Garmage geometry images using bilinear interpolation, resulting in a fine-grained initial draping state. 
In contrast to existing garment modeling frameworks that typically rely on coarse rigid transformations to position each panel, our Garmage-based approach provides vertex-level precision in the initial 3D placement. This capability allows us to accurately capture intricate folding behaviors and nuanced garment structures.

\section{GarmageSet}
\label{sec:garmageset}

As noted, Garmage’s vertex-level sewing and precise 3D initialization excel at modeling intricate drapes and folds, whereas existing datasets~\cite{korosteleva2024garmentcodedata,luo2024garverselod,zhu2020deep} are restricted to simple, flat garments and cannot fully evaluate our framework. To address this gap, we assembled a professionally curated, industrial-grade dataset \emph{GarmageSet} showcasing complex folding behaviors and multi-layer structures, complete with manually validated structural and style annotations, as well as multimodal augmentations including line-art sketches and sampled point clouds.

\subsection{GarmageSet Construction}  
GarmageSet comprises \(N=14{,}801\) unique garments spanning five major clothing categories: tops, pants, skirts, dresses, and outerwear. All garments are draped onto an A-posed standard avatar\footnote{Size S mannequin with Asian size 84.} to diminish the geometric variance brought by body sizes and poses.

\subsubsection{Data Acquisition} 

Building \emph{GarmageSet} entirely by hand would be prohibitively time-consuming. To scale the dataset construction efficiently, we adopt a component-centric strategy inspired by GarmentCode~\cite{korosteleva2023garmentcode}. As illustrated in Figure~\ref{fig:data_prep}, we first construct a structured component library from in-the-wild sewing patterns and then task professional modelers with assembling garments by randomly selecting components, applying design modifications (e.g., adjusting width or length, or adding decorative features), and combining them into complete garments.

To build the component library, we collect a diverse set of raw sewing patterns and engage professional pattern makers to annotate them following the hierarchical garment structure definitions detailed in Section~\ref{sec:garment_structure}. This process yields a well-organized collection of reusable garment parts, categorized by role (e.g., bodice, sleeve, collar) and tagged with stylistic attributes curated by experienced fashion designers and pattern makers\footnote{Some style tags and illustrations are adapted from Fashionpedia~\cite{fashionpedia}.}, as shown in Figure~\ref{fig:data_design_space}.

We then randomly sample valid combinations of components from the library and use QWen3 to propose $1$–$3$ modification instructions for each combination, such as altering silhouette proportions or adding style-specific elements. These modified configurations are assigned to professional garment modelers, who manually implement the design changes and assemble the components into finalized 3D garments.

\subsubsection{Garment Structure Definition}
\label{sec:garment_structure}
\begin{figure}[!ht]
    \centering
    \includegraphics[width=\linewidth]{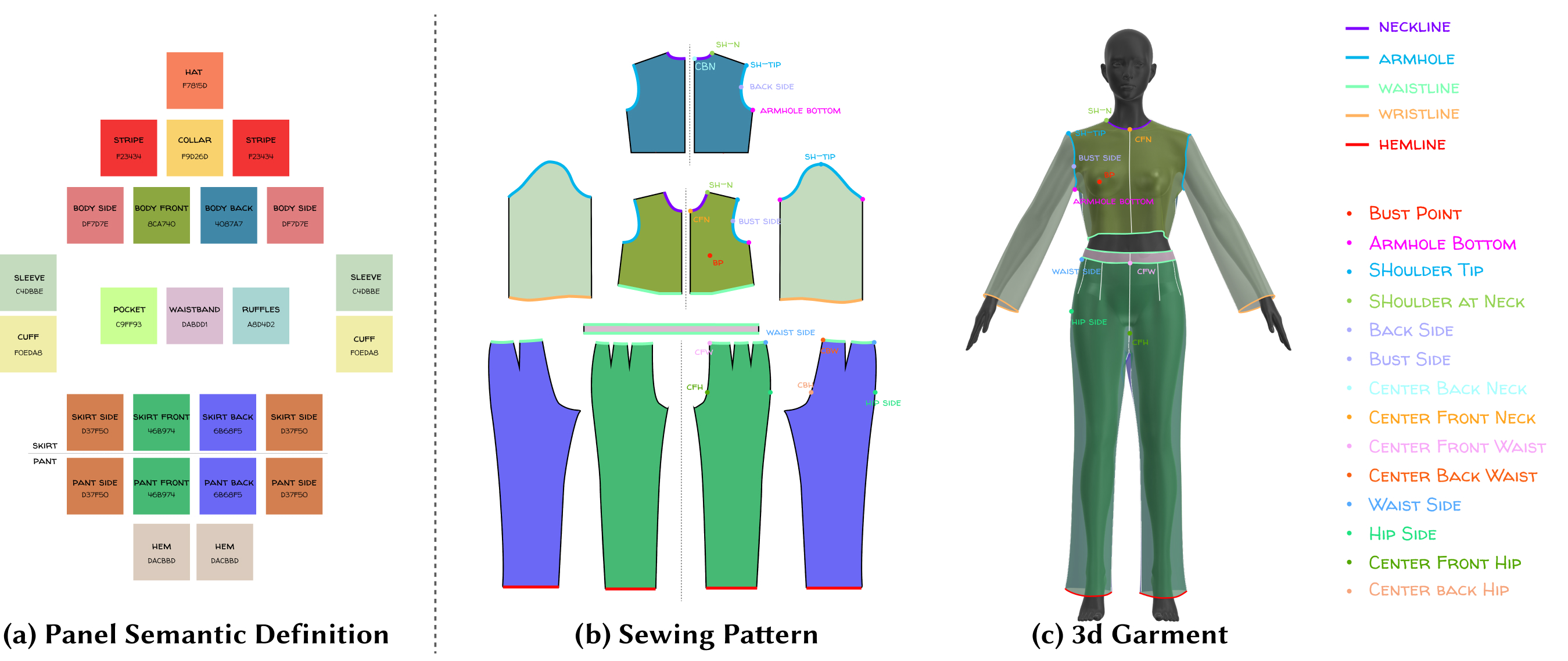}
    \caption{Garment structure definition and corresponding visualization on both sewing pattern space and 3D garments. (a) Color-coded definitions of eight structural (e.g., body front, sleeve) and seven decorative (e.g., pocket, ruffle) panel classes. (b) A sewing-pattern layout annotated by these semantic labels. (c) The corresponding 3D draped garment on the standard avatar, with each panel rendered according to its semantic class.
}
    \label{fig:garment_structure}
\end{figure}

As illustrated in Figure~\ref{fig:garment_structure}, garments exhibit a hierarchical structure comprising panels, edges, and landmarks, each capturing distinct semantic and geometric characteristics essential for garment design and construction. To accurately represent and leverage these hierarchical details, we introduce a structured annotation scheme that clearly defines panel-level semantics, structural lines, and fashion landmarks, as described below.

\begin{table}[!ht]
\caption{Panel counts ($\#$Panels) and mean average precision (AP) for semantic segmentation by our fine‐tuned PointTransformer v3, used to derive point‐cloud embeddings for conditional Garmage synthesis. The uniformly high AP values across all categories confirm the model’s robustness in extracting panel‐level semantics from unstructured point clouds, thereby providing a reliable conditioning signal.}
\label{tab:per_category_AP}
\resizebox{\linewidth}{!}{%
\begin{tabular}{c|ccccc}
\toprule
\rowcolor[rgb]{0.33,0.33,0.33}\textcolor{white}{Category} & \textcolor[HTML]{F9D26D}{collar}       & \textcolor[HTML]{C4DBBE}{sleeve}    & \textcolor[HTML]{8CA740}{body front} & \textcolor[HTML]{4087A7}{body back}    & \textcolor[HTML]{D37F50}{body side} \\ \midrule
\texttt{\#}Panels       & 9807     & 22576 & \textbf{34138}   & 22608   & 2857   \\ 
AP       & 0.95    & 0.97 & 0.94   & 0.94   & 0.40   \\ 
\midrule
\rowcolor[rgb]{0.33,0.33,0.33}\textcolor{white}{Category} & \textcolor[HTML]{46B974}{skirt/pant front}  & \textcolor[HTML]{6B68F5}{skirt/pant back}        & \textcolor[HTML]{B99946}{skirt/pant side} & \textcolor[HTML]{F7815D}{hat} & \textcolor[HTML]{F23434}{stripe}   \\ \midrule
\texttt{\#}Panels       & 28782    & 25242      & 3491      & 2965     & 1142      \\ 
        AP           & 0.93    &  0.95   & 0.40    & 0.98   & 0.69    \\ 
\midrule
\rowcolor[rgb]{0.33,0.33,0.33}\textcolor{white}{Category}  & \textcolor[HTML]{F0EDA8}{cuff}    & \textcolor[HTML]{DABDD1}{waist} & \textcolor[HTML]{DACBBD}{hem}       & \textcolor[HTML]{C9FF93}{pocket} & \textcolor[HTML]{A8D4D2}{ruffles}   \\ \midrule
\texttt{\#}Panels       & 9509   & 16880   & 3088      &  15571    & 2498     \\ 
        AP          & \emph{0.98}  & 0.96  & 0.79    & 0.93    & 0.42    \\ 
\bottomrule
\end{tabular}%
}
\end{table}

\emph{Panel-level semantics} are established by professional pattern makers based on panel shape, functional role, and placement relative to the human body. As shown in Figure~\ref{fig:garment_structure}, we identify eight structural classes—\textcolor[HTML]{F9D26D}{collar}, \textcolor[HTML]{C4DBBE}{sleeve}, \textcolor[HTML]{8CA740}{body front}, \textcolor[HTML]{4087A7}{body back}, \textcolor[HTML]{D37F50}{body side}, \textcolor[HTML]{46B974}{skirt/pant front}, \textcolor[HTML]{6B68F5}{skirt/pant back}, and \textcolor[HTML]{B99946}{skirt/pant side}—as well as seven decorative classes—\textcolor[HTML]{F7815D}{hat}, \textcolor[HTML]{F23434}{stripe}, \textcolor[HTML]{F0EDA8}{cuff}, \textcolor[HTML]{DABDD1}{waist}, \textcolor[HTML]{DACBBD}{hem}, \textcolor[HTML]{C9FF93}{pocket}, and \textcolor[HTML]{A8D4D2}{ruffles}. Annotators assign these semantic labels directly on the 2D sewing patterns using a customized LabelStudio~\cite{LabelStudio} annotation tool. Table~\ref{tab:per_category_AP} reports per-category panel counts in GarmageSet and the average precision (AP) of a panel classifier trained on it. Meanwhile, we use these panel-level semantics to fine-tune a Point Transformer v3 as the point-cloud feature encoder, providing semantic priors for point-cloud–conditioned Garmage generation.

Utilizing per-panel semantic annotations, we extract five types of \emph{structural lines} that define interfaces between semantic panel groups. The \textcolor[HTML]{8000FF}{neckline} delineates boundaries between front/back bodice and collar panels; \textcolor[HTML]{00B5EB}{armholes} separate bodice panels from sleeve panels; \textcolor[HTML]{80FFB4}{waistline} defines the interface between waist panels and adjacent bodice or skirt/pant panels (or directly between bodice and skirt/pant panels if waist panels are absent); \textcolor[HTML]{FFB360}{wristline} marks the junction between sleeves and cuffs or the lower edge of sleeves if cuffs are absent; and \textcolor[HTML]{FF0000}{hemline} represents the boundary between bodice/skirt/pant panels and hem panels, or the lower edge of these panels when hem panels do not exist. During training, we augment our dataset by perturbing these structural lines, simulating realistic variations in sleeve length, garment length, waist height, and other key design parameters.

\begin{figure*}[!t]
    \centering
    \includegraphics[width=\linewidth]{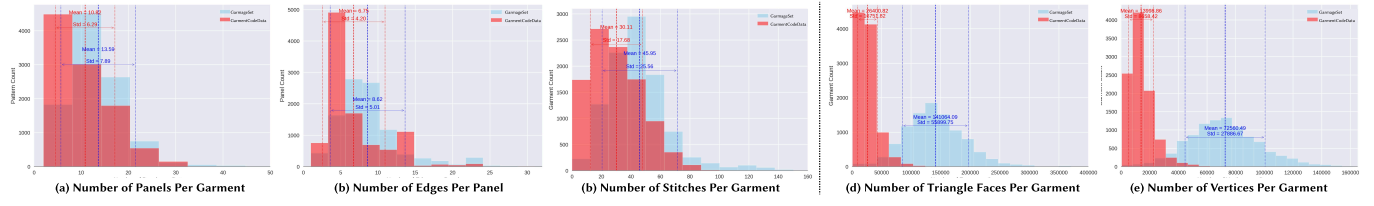}
    \caption{Dataset statistics comparing \textcolor[HTML]{B8DBEE}{\emph{GarmageSet}} and \textcolor[HTML]{F47478}{\emph{GarmentCodeData}}~\cite{korosteleva2024garmentcodedata}. Histograms illustrate (a) panels per garment, (b) edges per panel, (c)  per garment, (d) mesh vertices per garment, and (e) mesh faces per garment distribution among the $10,000$ sampled garments (or $10,000$ panels) from both datasets. Dashed lines indicate the mean and standard deviation for each distribution. GarmageSet exhibits higher average values and broader variance across all metrics, indicating enhanced structural complexity and superior drape fidelity.}
    \label{fig:data_stat}
\end{figure*}

\emph{Fashion landmarks} serve as critical reference points for precise garment construction and fitting. Examples include the \textcolor[HTML]{06B3F1}{shoulder tip (SH)}, \textcolor[HTML]{FF2500}{bust point (BP)}, \textcolor[HTML]{FFA51F}{center front neck (CFN)}, and \textcolor[HTML]{F8AAF8}{center front waist (CFW)}. These landmarks are annotated on both the 2D sewing patterns and their corresponding 3D models using consistent vertex IDs on the mesh. Such dual annotations help align sewing patterns from different garments into a standardized 2D space, eliminating positional ambiguity and facilitating more effective learning. Additionally, these landmarks are consistently projected onto multi-view 2D images, significantly enriching existing fashion landmark datasets and improving the accuracy of fashion landmark estimation and retrieval models, ultimately offering comprehensive support for diverse fashion AI applications.

\subsubsection{Data Formation And Multi-modal Augmentation}
\label{sec:data_formation}
For each garment, we partition its raw 2D patterns into individual panels \(P_i\) and compute their physical dimensions \(\mathbf{d}_i\) and axis-aligned bounding boxes \(\mathbf{B}_i\).  We then rasterize each cloth piece’s normalized 3D mesh \(C_i\) into a \(256\times256\times4\) geometry image \(I_i\), and construct the Garmage representation for the garment (Section~\ref{sec:garmage-repr}).

The ground-truth \emph{sewing information} is stored as vertex–vertex pairs $(v_a^i, v_b^j)$ from cloth pieces $C_i$ and $C_j$. During Garmage rasterization, each vertex $v_a^i$ is projected to a 2D pixel coordinate $u_a^i\in[0,1]^2$. We encode panel identity by shifting the integer part of the pixel coordinate with the panel index, and thus reformulate seams in a panel–pixel domain of shape $M\times 2\times 2$:
\begin{equation}
\begin{split}
   &\mathcal{S}=\{s_k\}_{k=1}^M\in\mathbb{R}^{M\times 2\times 2},\;\; \text{ where}\\
   &s_k = (i+u_a^i,\,j+u_b^j) = \bigl(\emph{Rast}(v_a^i,C_i),\,\emph{Rast}(v_b^j,C_j)\bigr).
\end{split}
\end{equation}

To train GarmageNet under \emph{multi-modal conditions}, we align each Garmage \(\mathcal{G}\) with four modalities: 

\begin{itemize}
    \item A manually annotated \emph{short sentence} captures each garment’s category, silhouette, and design details according to a set of professionally defined dimensions (Figure ~\ref{fig:data_design_space}). During modeling, we ask the designers to label all applicable dimensions for a given garment asset and leverage Qwen3~\cite{qwen3} to reformat the annotation as a CLIP-compatible, comma-separated string, with the first segment always denoting the garment category. During training, we randomly delete at most $4$ design detail descriptions.

    \item A set of \emph{line-art sketches and clay renderings} to capture each garment’s visual characteristics. These images are rendered from $24$ uniformly sampled camera viewpoints arranged on a circle centered on the garment. The circle’s radius is automatically adjusted so that, in the frontal view, the garment could nearly fill the frame. All sketches and clay renderings are output at $3840\times 2048$ resolution, and we record each camera’s transformation matrix in the standard NeRF format.

    \item A \emph{point cloud} sampled from the garment mesh using Poisson‐disk sampling (Open3D) to capture its geometric detail. To closely mimic real-world scans or multi-view reconstructions, which emphasize the exterior surface, we adapt sampling density by occlusion: outer panels are sampled at a high density, while inner panels use a sparser density. We randomly downsample these point clouds at varying rates to improve model robustness and performance.    
\end{itemize}

\subsection{Dataset Statistics}  
As summarized before, GarmageSet comprises $14{,}801$ professionally modeled garments across five major categories including tops (2,888), outerwears (2,293), pants (857), skirts (1,523), dresses (6,454), and 786 garments from other categories such as sportswear, bras, pajamas and cheongsams. 
Each garment is annotated along professionally defined design dimensions with over a hundred part‐wise variations (Figure ~\ref{fig:data_design_space}), yielding a combinatorial design space of more than \(2.9454\times10^{11}\) topologically distinct configurations. 

\begin{figure*}[!htp]
    \centering
    \includegraphics[width=\linewidth]{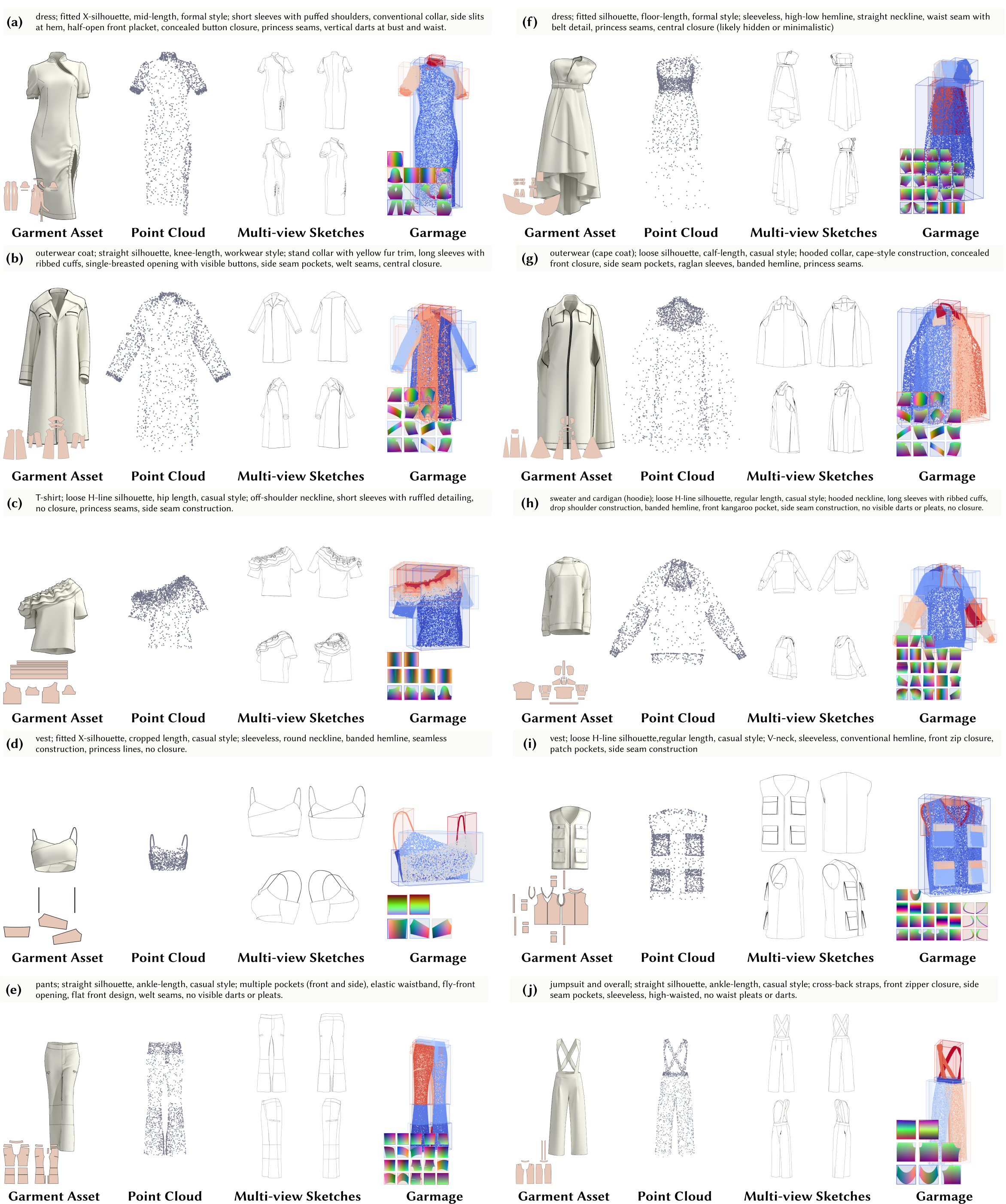}
    \caption{Representative examples from our \emph{GarmageSet}, demonstrating the dataset's rich diversity in garment categories, styles, and intricate folding patterns.}
    \label{fig:data_example}
\end{figure*}

Although smaller in size, GarmageSet covers substantially richer variation than GarmentCodeData~\cite{korosteleva2024garmentcodedata}, which is limited to basic modifications (e.g., a single dart type for \texttt{FittedShirt} and single lapel style defined in \texttt{SimpleLapel}). To quantify structural complexity, we randomly sample $10{,}000$ garments (and $10{,}000$ panels) from each dataset and compare statistics in Figure ~\ref{fig:data_stat}. GarmageSet has on average $13.59_{\pm7.89}$ panels and $46.01_{\pm26.45}$ stitches per garment with $8.62_{\pm5.01}$ edges per panel. By contrast, GarmentCodeData provides only $10.82_{\pm6.29}$ panels and $30.26_{\pm17.59}$ stitches per garment, with $6.75_{\pm4.20}$ edges per panel, indicating significantly lower structural richness.
In terms of 3D drape fidelity, by setting the particle distance to $6$mm during simulation, GarmageSet features \(72{,}560.5_{\pm27{,}886.7}\) vertices and \(141{,}064.1_{\pm55{,}899.8}\) faces per garment; while GarmentCodeData only has $13{,}998.86_{\pm 8{,}658.42}$ vertices and $26{,}400.82_{\pm16{,}751.82}$ faces per garment, demonstrating that GarmageSet delivers over five-fold higher mesh resolution than GarmentCodeData which indicates richer structural details.
Figure ~\ref{fig:data_example} presents representative samples from GarmageSet, visually demonstrating its high geometric fidelity and intricate structural detail. For example, complex garment foldings and shirrings (c,f); multi-layered design (d) and irregular splits (a,i,e,h) that hard to achieve with GarmentCode's parametric formulation.

\section{Experiments}
\label{sec:experiment}

In this section, we first detail the implementation and training protocols for \textit{GarmageNet} and \textit{GarmageJigsaw} then quantify our frameworks’ performance by evaluating sewing‐pattern recovery quality, 3D geometry fidelity, and sewing accuracy.

\subsection{Implementation Details}
We randomly reserved 1,024 garment assets from \textit{GarmageSet} for validation, using the remaining 13,777 assets for training.

GarmageNet was trained on a single NVIDIA A100 GPU over 1–2 days using a two‐stage protocol. In the latent‐encoding stage, we trained the VAE for 200 epochs with a batch size of 256, using the AdamW optimizer at a learning rate of \(5\times10^{-4}\). This stage completes in approximately 2 hours.
In the diffusion‐generation stage, we employ a standard DDPM scheduler and train the denoiser for 20,000 epochs with a batch size of 4,096, which takes approximately 12 hours. 
In conditional generation with text prompts or point clouds, we need to incorporate augmentations such as random word dropout in prompts, variable point‐cloud sampling densities, and on‐the‐fly embedding computation. Thus, extends total training time to roughly 24 hours.

We trained GarmageJigsaw using two NVIDIA RTX 4090 GPUs with a batch size of 28. The training was initialized with a learning rate of $1\times10^{-3}$, which was gradually decreased using cosine learning rate decay, ultimately reaching $2\times10^{-5}$ at the end of the training process. We train our \textit{GarmageJigsaw} for $100$ epochs, taking approximately 27 hours in total.

\subsection{Evaluation And Comparison}
As previously demonstrated, GarmageNet can synthesize complete garment assets, encompassing 2D sewing patterns, sewing correspondences, and high‐resolution 3D initializations. Accordingly, we evaluate its generation quality across these core dimensions.

\subsubsection{Cross-Dataset Comparison on Sewing Pattern Quality}
First, we conduct a cross-dataset study to quantify the impact of GarmageSet (real production data) on model performance and to validate GarmageNet against a representative \textit{structure-centric} baseline. We compare with AIpparel~\cite{nakayama2024aipparel}, which quantizes sewing patterns into 1D token sequences and uses a large language model to decode patterns from multimodal inputs. Evaluations are performed on the synthetic GCD-MM dataset used in AIpparel and on our GarmageSet. 
Because Garmage adopts a rasterized sewing-pattern representation that differs fundamentally from vector-quantized encodings, and because GarmageSet both violates several simplifying assumptions of GCD-MM (e.g., one-to-one edge-wise seams) and lacks ground-truth seam graphs and intermediate initialization cues such as per-panel translations/rotations (Section~\ref{sec:related_garmet_dataset_sp}), AIpparel’s original metrics, such as Panel-L2, edge-wise Stitch-F1, Trans-L2, Rot-L2, are not directly applicable. We therefore introduce the following metrics that are valid on both GCD-MM and GarmageSet to ensure a fair comparison:

\begin{table}[t]\centering
\caption{Cross-dataset comparison of GarmageNet with vector-quantization-based representative AIpparel~\cite{nakayama2024aipparel} on GarmageSet and GCD-MM. GarmageNet generally outperforms AIpparel on per-panel intersection over union (Panel-IoU), simulation succession rate (SSR) and chamfer distance (CD), but slightly lags on predicted number of panels (\#Panels) and stitch‑F1 score.}
\label{tab:cross_dataset_comp}
\scriptsize
\resizebox{\linewidth}{!}{%
\begin{tabular}{cllcccccc}\toprule
\textbf{Exp.} & \textbf{Dataset} &\textbf{Method} &\textbf{Panel-IOU} &\textbf{\#Panels} &\textbf{Stitch-F1} &\textbf{SSR} &\textbf{CD (mm)} \\\midrule
1 & \multirow{3}{*}{GCD-MM} &AIpparel &0.89 &\cellcolor[HTML]{f1f0fc}\textbf{0.85} &0.76 &0.59 &59.3 \\
2 & &AIpparel-Rand &0.43 &0.26 &0.17 &/ &/ \\
3 & &GarmageNet &\cellcolor[HTML]{f1f0fc}\textbf{0.94} &0.47 &\cellcolor[HTML]{f1f0fc}\textbf{0.66} &\cellcolor[HTML]{f1f0fc}\textbf{0.52} &\cellcolor[HTML]{f1f0fc}\textbf{30.1} \\\midrule
4 & \multirow{2}{*}{GarmageSet} &AIpparel &0.53 &0.16 &/ &/ &/ \\
5 & &GarmageNet &\cellcolor[HTML]{f1f0fc}\textbf{0.88} &\cellcolor[HTML]{f1f0fc}\textbf{0.23} &\cellcolor[HTML]{f1f0fc}\textbf{0.78} &\cellcolor[HTML]{f1f0fc}\textbf{0.91} &53.94 \\
\bottomrule
\end{tabular}}
\end{table}

\begin{itemize}
    \item \textbf{Panel Shape Complexity.} We use \textit{pixel-wise IoU} between rasterized panels, which is able to capture internal holes, multi-edge loops, and complex contours (e.g., spirals) present in GarmageSet.
    \item \textbf{Sewing Quality.} GarmentCodeData encodes seams as \textit{one-to-one, full-edge} correspondences, whereas GarmageSet requires \textit{many-to-many, partial-edge} mappings for multilayering and pleats (enforcing 1:1 full-edge sewing can explode edge counts). In the absence of a robust sewing-quantization for GarmageSet, we report \textit{Stitch-F1} where labels are compatible—namely on GarmentCodeData and GarmageNet outputs (Exps.~1–3,5).
    \item \textbf{3D Initialization Cues.} As noted in Sec.~\ref{sec:related_garmet_dataset}, GarmageSet lacks per-panel rigid placements (translations/rotations), rendering AIpparel’s Trans-L2 and Rot-L2 inapplicable. We therefore evaluate drape quality using Simulation-Success Rate (SSR) and Chamfer Distance (CD) between successfully simulated predictions and ground-truth meshes.
\end{itemize}

We retrained GarmageNet on GarmentCodeData for 50 GPU-hours (RTX 4090) and AIpparel on GarmageSet for 157 GPU-hours (H20). All metrics are reported on line-art–guided generation over test splits of corresponding datasets. Table~\ref{tab:cross_dataset_comp} lists the detailed comparison results\footnote{Exp.1 is evaluated with AIpparel's~\cite{nakayama2024aipparel} officially provided checkpoints.}, comparing Exp.1 vs Exp.3, Exp.4 vs Exp.5, we can conclude that GarmageNet generally outperforms AIpparel in Panel-IoU, SSR, and CD, but trails slightly in \#Panels and Stitch-F1 on the GCD-MM dataset. We attribute this degradation in part to the sparsity of line-art features and the presence of occlusions, which can cause the pretrained feature extractor to miss fine segmentation details critical to \#Panels or to overlook inner panels that are partially hidden (Figure~\ref{fig:limitations} (e,f)).

\begin{figure*}[!t]
    \centering
    \includegraphics[width=\linewidth]{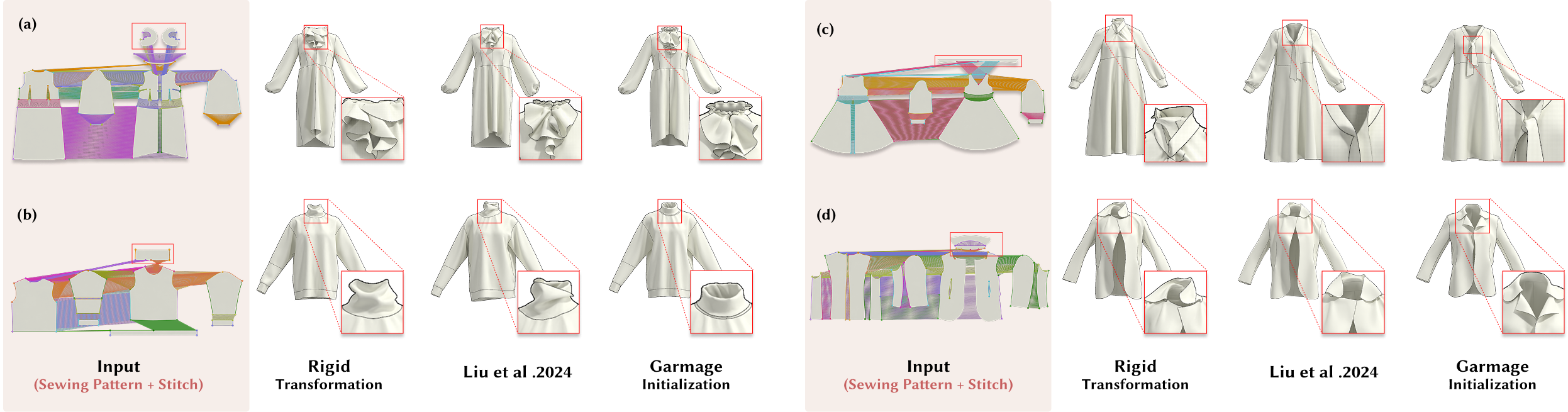}
    \caption{Qualitative comparison between Garmage, rigid-transformation-based and optimization-based~\cite{liu2024automatic} garment initialization methods. Garmage effectively models intricate folding structures such as Jabot decorations (a), folded-over turtlenecks (b), natural knots (c), and lapel structures (d).}
    \label{fig:eval_ssr}
\end{figure*}

\subsubsection{Sewing Accuracy Evaluation} 

The \textit{GarmageJigsaw} module for sewing recovery consists of two main components: a point classifier and a sewing predictor. To thoroughly evaluate its effectiveness, we assess each component independently and perform an ablation study on input feature selection, comparing the fused 2D/3D features used in GarmageJigsaw with 2D-only and 3D-only variants.

The point classifier operates as a binary classifier, and we evaluate its performance using precision and recall. The \textbf{classification precision (CP)} measures the proportion of correctly identified sewing points (i.e., true positives) among all predicted positives, while the \textbf{classification recall (CR)} indicates the proportion of true positive predictions among all sewing points in the ground truth. 
As shown in Table~\ref{tab:evaluate_garmagejigsaw}, the point classifier achieves a precision of $99.16\%$ and a recall of $97.13\%$, indicating strong performance in identifying sewing points.

\begin{table}[t]\centering
\caption{Ablation study on sewing relationship recovery, comparing the performance of GarmageJigsaw trained with both 2D and 3D features versus models trained with only 2D or 3D features. The table reports key metrics including point classification precision (CP), recall (CR), average matching distance (AMD), topological accuracy (tACC), and topological precision (tP).}
\label{tab:evaluate_garmagejigsaw}
\scriptsize
\begin{tabular}{lccccccc}
\toprule
& \textbf{CP} ($\uparrow$) & \textbf{CR} ($\uparrow$) & \textbf{AMD} ($\downarrow$) & \textbf{tACC} ($\uparrow$) & \textbf{tP} ($\uparrow$)  \\
\midrule
GarmageJigsaw & 99.16 & \cellcolor[HTML]{f1f0fc}\textbf{97.13} & \cellcolor[HTML]{f1f0fc}\textbf{6.610} & \cellcolor[HTML]{f1f0fc}\textbf{96.79} & \cellcolor[HTML]{f1f0fc}\textbf{98.68}  \\
3D-feat Only & \cellcolor[HTML]{f1f0fc}\textbf{99.27} & 96.99 & 7.790 & 96.36 & 97.96  \\
2D-feat Only & 99.21 & 97.12 & 10.59 & 96.28 & 97.70  \\
\bottomrule
\end{tabular}
\end{table}

For the sewing predictor, we first evaluate the \textbf{panel-level} topological quality of the generated sewing patterns with:
\begin{itemize}
    \item \textbf{Accuracy (tACC)}: The proportion of correctly predicted sewing connections (correct sewing pairs) out of all predicted connections. Higher values indicate better topological correctness.
    \item \textbf{Precision (tP)}: The proportion of correctly predicted sewing connections out of all predicted connections, where higher values reflect fewer false positives.
\end{itemize}

Additionally, we evaluate vertex-level sewing quality using \textbf{Average Matching Distance (AMD)}, which calculates the average Euclidean distance between predicted sewing correspondent and ground truth correspondent for all vertices. Lower AMD values indicate better alignment between predicted and actual sewing positions.

Table~\ref{tab:evaluate_garmagejigsaw} summarizes the evaluation results with ablation studies on using only 2D or 3D features for sewing relationship recovery. These results confirm that combining both 3D and 2D features enables GarmageJigsaw to achieve more robust stitching recovery with lower AMD value and topological accuracy.

\subsubsection{Fine‐Grained 3D Initialization Evaluation}
To quantify the benefits of GarmageNet’s vertex‐level initializations, we compare its simulation succession rate (SSR) against two baselines: (1) rigid transformations-based initialization as used in GarmentCodeData~\cite{korosteleva2023garmentcode}; and (2) optimization‐based initialization from raw sewing patterns~\cite{liu2024automatic}. 

We collect 150 sewing patterns with ground truth stitching relationships from GarmageSet, recover their initial drape status with GarmentNet and drape onto our standard Size S avatar using identical simulation settings as ~\cite{liu2024automatic} and compare the SSR as garments draped successfully onto the avatar without observable self-collision, body-collision, sliding errors \textit{etc.} 
For the rigid baseline, we leverage the per‐panel semantics in GarmageSet (Section~\ref{sec:data_formation}) to assign each panel a fixed pose, using standard rigid placement for body, skirt, and front/back panels, and cylindrical arrangement for tubular components such as sleeves and collars.

As a result, GarmageNet achieves an SSR of 91.41\%, substantially higher than rigid initialization (59.38\%) and on par with optimization-based initialization (93.75\%).
Figure~\ref{fig:eval_ssr} presents representative cases, from which we can conclude that our fine-grained, per-vertex placements could provide robust draping initialization for complex draping (a), knots (c) and folding behaviors (b,d).

\begin{figure*}[!t]
    \centering
    \includegraphics[width=\linewidth]{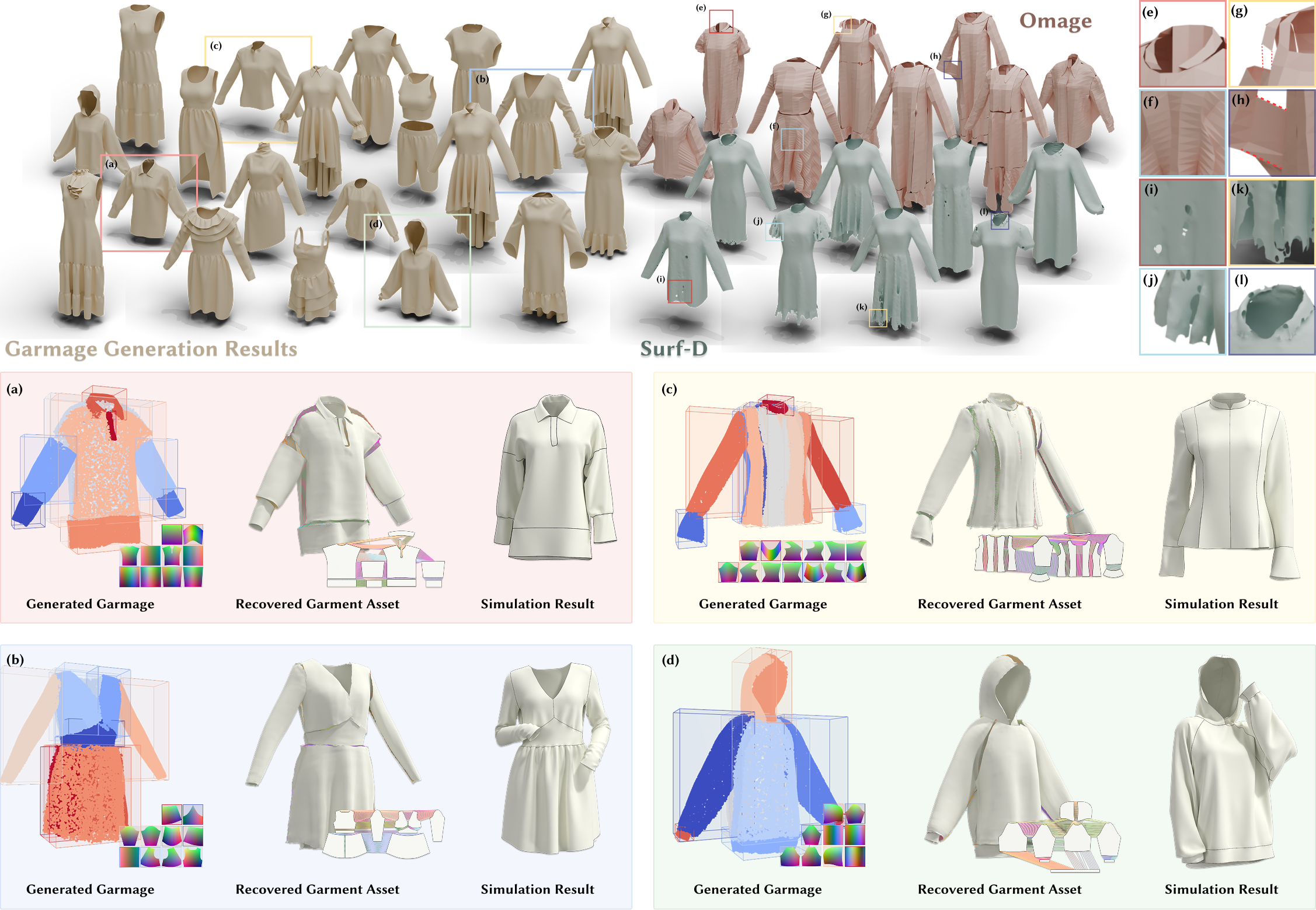}
    \caption{Unconditional garment generation comparison between \textbf{GarmageNet}, \textbf{Omage}~\cite{yan2024object}, and \textbf{Surf-D}~\cite{yu2025surf}. GarmageNet (left block) produces simulation-ready assets complete with vectorized sewing patterns, vertex-wise stitch relationships, and fine-grained 3D draping initializations (a,b,c,d). In contrast, Omage’s outputs (top right) exhibit incomplete panels (g), grid-like tessellation artifacts (f), erroneous stitching between non-adjacent panels (h), and spurious triangles that connect a panel’s boundary vertices back to the global origin (e). Surf-D’s meshes (bottom right) suffer from unwanted holes (i, l) and frayed, irregular boundaries (j, k). These close-up comparisons highlight GarmageNet’s superior geometric fidelity, coherent panel topology, and artifact-free mesh integrity.}
    \label{fig:uncond_gen}
\end{figure*}

\begin{table}[!t]\centering
\caption{Comparison of generation quality, diversity, and efficiency between \textit{GarmageNet}, \textit{Omage}\cite{yan2024object}, and \textit{Surf-D}\cite{yu2025surf}. Quality metrics include Minimum Matching Distance (MMD, $\times10^{-3}m$), Jensen–Shannon Divergence (JSD), point-cloud FID (p-FID), and point-cloud KID (p-KID), where lower values indicate better fidelity. Diversity is measured by Coverage (COV, \%), and efficiency is assessed based on inference GPU memory usage (Mem.) and inference speed (measured in seconds).}\label{tab:quality_comp}
\scriptsize
\resizebox{\linewidth}{!}{%
\begin{tabular}{l|cccc|c|cc}\toprule
\multirow{2}{*}{\textbf{Method}} &\multicolumn{4}{c|}{\textbf{Quality}} &\textbf{Diversity} &\multicolumn{2}{c}{\textbf{Efficiency}} \\
&MMD ($\downarrow$) &JSD ($\downarrow$) & p-FID ($\downarrow$) & p-KID ($\downarrow$) &COV ($\uparrow$) &Mem. ($\downarrow$) &Duration ($\downarrow$)\\\toprule
Surf-D &146.87 &0.7907 &46.61 &0.1718 &16.02\% & 7 GB & 25.7s \\
Omages &96.44 &0.1185 &29.38 &0.1271 &28.16\%  & 3.3 GB & 120s \\
Ours &\cellcolor[HTML]{f1f0fc}\textbf{35.55} & \cellcolor[HTML]{f1f0fc}\textbf{0.0337} &\cellcolor[HTML]{f1f0fc}\textbf{15.34} & \cellcolor[HTML]{f1f0fc}\textbf{0.029} &\cellcolor[HTML]{f1f0fc}\textbf{41.02\%}  & 4 GB & 8s\\
\bottomrule
\end{tabular}}
\end{table}

\subsubsection{3D Garment Asset Quality}

We compare GarmageNet’s garment generation quality against two representative \textit{geometry-centric} generation approaches: Omage~\cite{yan2024object}, which typifies geometry-image-based 3D generation pipelines akin to our approach; and Surf-D~\cite{yu2025surf}, an implicit-field method that generates surfaces via unsigned distance functions.
Table~\ref{tab:quality_comp} presents the comparison results according to five metrics:
\begin{itemize}[nosep]
\item \textbf{Minimum Matching Distance (MMD)} measures the average closest‐distance between each real sample and its generated counterpart (units of $10^{-3}m$). A lower MMD indicates that, on average, every real garment has a very similar counterpart among the generated set.
\item \textbf{Jensen–Shannon Divergence (JSD)} quantifies the overall distributional discrepancy. A lower JSD means that the probability distributions of real and generated samples are more similar. 
\item \textbf{Point-cloud FID (p-FID)} and \textbf{KID (p-KID)} assess generation fidelity using learned feature embeddings, with lower values indicating the generated feature distribution are closer to those of the real data.
\item \textbf{Coverage (COV)} is the fraction of real samples matched by at least one generated sample (in percentage $\%$). A higher COV indicates broader exploration of the real data manifold, i.e., greater diversity.
\end{itemize}
For a fair comparison, all baseline methods were retrained on the full GarmageSet under the unconditional generation setting. Specifically, Omage~\cite{yan2024object} was trained at a resolution of $64 \times 64$, requiring approximately 50 hours for training and consuming 3.3MB of memory with an inference time of 120 seconds per sample. Surf-D~\cite{yu2025surf} was trained at a resolution of $512$, where the VAE module took four days to train on two RTX 4090 GPUs, followed by 20 hours of diffusion model training. To compute point-cloud FID and KID scores, we adopt the pretrained PointNet++ feature extractor provided by Point-E~\cite{nichol2022point}. Each method generated 128 random samples using a single NVIDIA GeForce RTX 3060 for evaluation. As reported in Table~\ref{tab:quality_comp}, GarmageNet outperforms both Omage and Surf-D in terms of generation fidelity, diversity, and computational efficiency.

Figure~\ref{fig:uncond_gen} presents unconditional generation results from GarmageNet alongside those of Surf-D and Omage. Omage produces a single multi-chart geometry image for the entire garment, making its outputs vulnerable to irregular UV chart packing; addressing this requires a much larger network and longer training times. As shown, Omage’s results appear coarse and often suffer from missing panels. Surf-D exemplifies a backward modeling approach, using an unsigned distance field (UDF) for generation and then extracting a triangle mesh. Consequently, it generates only a single, monolithic mesh without any explicit sewing-pattern structure, and the UDF-to-mesh conversion can introduce holes. In contrast, GarmageNet delivers panel-aware garments with complete and crisp per-panel structure, and fine-grained draping status.
\begin{figure*}[!t]
    \centering
    \includegraphics[width=\linewidth]{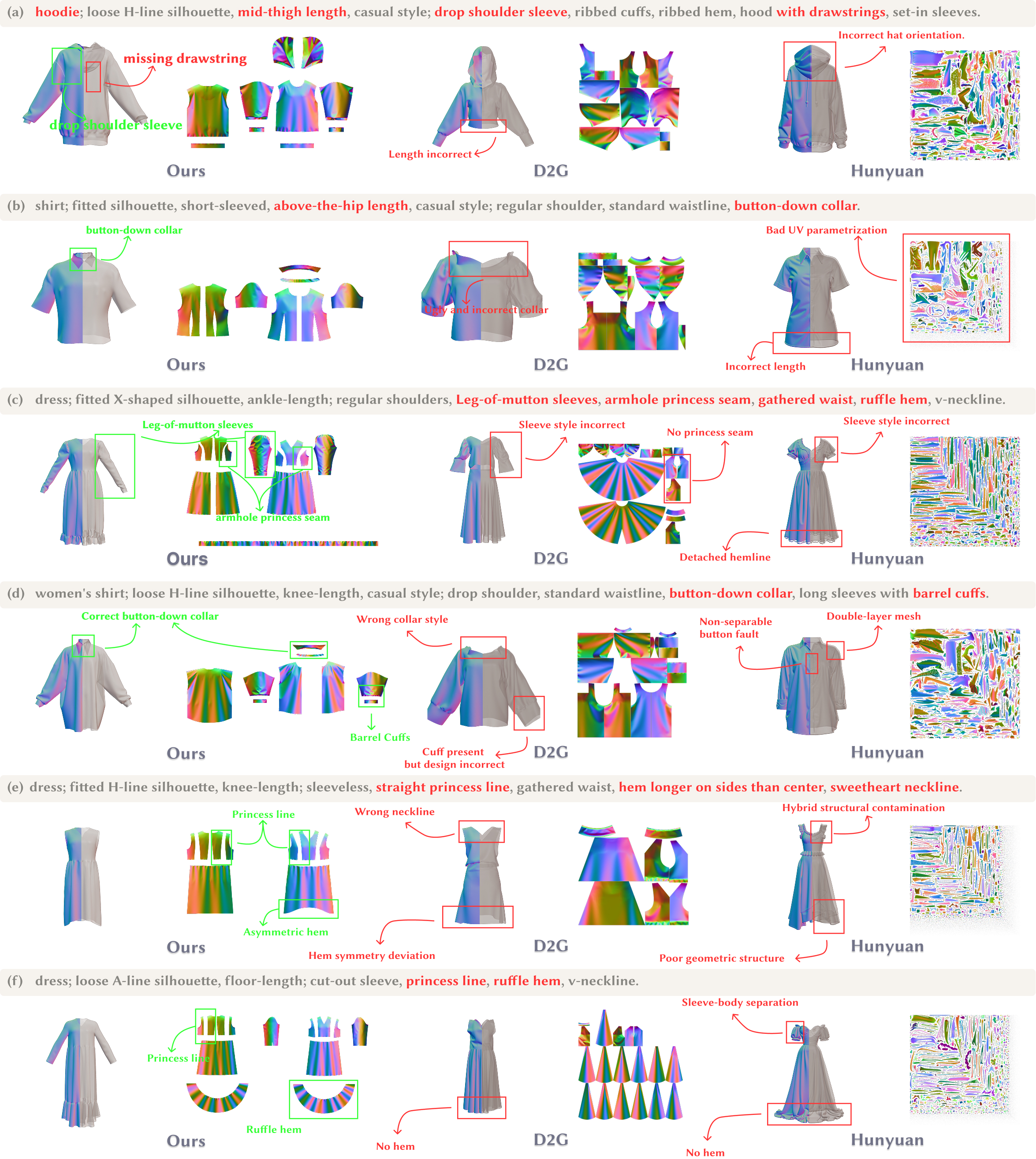}
    \caption{Text conditioned garment generation results and comparison with Design2GarmentCode~\cite{zhou2024design2garmentcode} and Hunyuan 3D 2.5~\cite{zhao2025hunyuan3d}.}
    \Description{Comparison with state-of-the-art 3D generation models}
    \label{fig:cond_gen_text}
\end{figure*}

\begin{figure*}[!t]
    \centering
    \includegraphics[width=\linewidth]{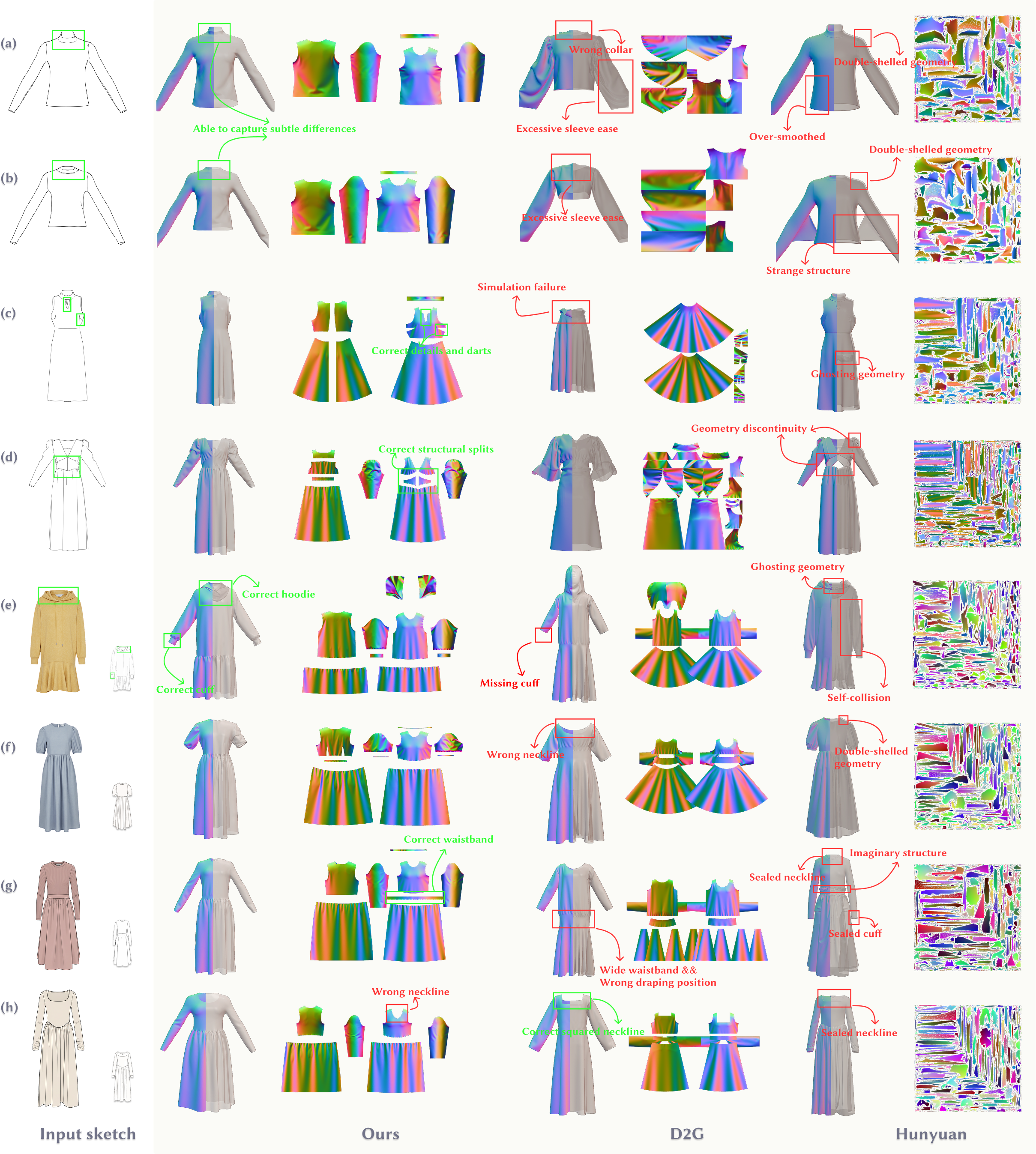}
    \caption{Line-art guided garment generation results and comparison with Design2GarmentCode~\cite{zhou2024design2garmentcode} and Hunyuan 3D 2.5~\cite{zhao2025hunyuan3d}.}
    \Description{Comparison with state-of-the-art 3D generation models}
    \label{fig:cond_gen_sketch}
\end{figure*}

\section{Applications}
\label{sec:application}

We demonstrate the practical versatility and effectiveness of the proposed \emph{GarmageNet} framework through four application scenarios that cover the full spectrum of digital garment modeling. These include interpreting abstract design concepts, automatically generating 3D garment assets from raw sewing patterns, reconstructing manufacturable sewing patterns from unstructured data, and performing conventional garment asset editing based on simple textual inputs. These scenarios showcase GarmageNet’s ability to accurately translate diverse inputs into structurally sound and visually compelling garment assets, bridging the gap between creative ideation and real-world garment production.

\begin{table*}[!t]\centering
\caption{User study results for generation quality comparison of our method against state-of-the-art (SOTA) forward generation technique Design2GarmentCode (trained on GarmentCodeData), and backward generation technique Hunyuan3D 2.5. Here, \textit{Agreement} evaluates the alignment between the generated garment and the design input (text or line-art sketch). \textit{Garment  Aesthetic} evaluates the geometric quality of the generated 3D garment asset, while \textit{Sewing Pattern Aesthetic} evaluates the quality of the generated sewing pattern. We provide CLIPScore as an additional agreement evaluation for text-guided generation.}\label{tab:quality_comp2}
\scriptsize
\begin{tabular}{c|cccc|ccc}\toprule
\multirow{2}{*}{\textbf{Method}} &\multicolumn{4}{c|}{\textbf{Text-Guided Generation}} &\multicolumn{3}{c}{\textbf{Line-Art Guided Generation}} \\
&Agreement &Garment Aesthetic & Sewing Pattern Aesthetic &CLIPScore &Agreement &Garment Aesthetic & Sewing Pattern Aesthetic \\ \midrule
\textbf{GarmageNet + GarmageJigsaw} & \cellcolor[HTML]{f1f0fc}\textbf{62.50}\% &\cellcolor[HTML]{f1f0fc}\textbf{85.00\%} &\cellcolor[HTML]{f1f0fc}\textbf{90.42\%} &\cellcolor[HTML]{f1f0fc}\textbf{0.3076} & \cellcolor[HTML]{f1f0fc}\textbf{77.34}\% & \cellcolor[HTML]{f1f0fc}\textbf{68.75}\% & \cellcolor[HTML]{f1f0fc}\textbf{97.66}\% \\
Design2GarmentCode~\cite{zhou2024design2garmentcode} &4.17\% &7.92\% &9.58\% &0.2955 & 0.0\% & 10.16\% & 2.34\% \\
Hunyuan 3D v2.5~\cite{zhao2025hunyuan3d} & 33.33\% &7.08\% &0.00\% &0.3016 & 22.66\% & 21.09\% & 0.0\% \\
\bottomrule
\end{tabular}
\vspace{2pt}
\end{table*}

\subsection{Design Concept to Garment Generation}
\label{sec:app_cond_gen}
Generating garments directly from high-level design concepts, such as textual descriptions or minimalistic line-art sketches, significantly streamlines fashion design workflows, particularly in rapid prototyping and initial visualization stages. Unlike traditional methods that necessitate detailed technical specifications, GarmageNet interprets natural language prompts and simple sketches to automatically produce structurally correct and visually coherent 3D garments.

Qualitative evaluations supported by detailed X-ray renderings and UV-aligned normal maps reveal that GarmageNet effectively captures original design intents. The generated garments exhibit clearly defined seam structures, realistic draping, and well-articulated folds—key elements often compromised in outputs from existing frameworks such as Design2GarmentCode (forward generation) and Hunyuan3D v2.5 (backward generation).

Figure~\ref{fig:cond_gen_text}, \ref{fig:cond_gen_sketch} provide qualitative evaluations of garments generated from text prompts and line-art sketches, compared against two state-of-the-art baseline models: the forward generation approach, Design2GarmentCode\cite{zhou2024design2garmentcode}, trained on GarmentCodeData~\cite{korosteleva2024garmentcodedata}, and the backward generation method, Hunyuan3D v2.5~\cite{zhao2025hunyuan3d}, trained on massive 3D assets. 
For each generated garment, we present X-ray renderings to reveal the underlying geometric structures and UV-aligned normal maps to intuitively assess the quality of the generated sewing patterns and the detailed fold structures. Our outputs demonstrate clear and accurate seam structures, precise garment draping, and refined folds which are inadequately represented by the baseline methods.

Leveraging the line-art sketch-conditioned GarmageNet as a baseline, our framework could further enable in-the-wild image guided garment generation. Specifically, we employ a LoRA~\cite{hu2022lora} fine-tuned FLUX~\cite{flux2024} model to transfer real-world photographs and design sketches to GarmageSet style line-art sketches. These sketches subsequently guide the Garmage generation process, with results illustrated in Figure~\ref{fig:cond_gen_sketch} (e-h), underscoring the model’s enhanced versatility and real-world applicability.

A comprehensive user study involving 20 professional fashion designers, pattern makers, and 3D apparel modelers validated our findings quantitatively. Each participant is asked to review 48 garment generation results (24 text prompts and 24 sketches randomly sampled from $1000+$) and select which method’s result was best under three criteria: 
\begin{itemize}[nosep]
    \item \textbf{Agreement} with the input prompt (i.e. how well the 3D garment matches the described or drawn design);
    \item \textbf{Garment Aesthetic} (overall visual and geometric quality of the 3D garment model);
    \item \textbf{Sewing Pattern Aesthetic} (quality and plausibility of the underlying pattern structure, as evident in the model and its UV seams).
\end{itemize}
The aggregated preference results (normalized percentages of selections for each model) in Table~\ref{tab:quality_comp2} indicate GarmageNet significantly outperformed the baselines across all metrics, being preferred in over 60\% of cases for Agreement, 85\% for Garment Aesthetic, and approximately 90\% for Sewing Pattern Aesthetic in text-guided generation; and 77\% for Agreement. 68.75\%for Garment Aesthetic and 97.66\% for Sewing Pattern Aesthetic. Further, GarmageNet achieved the highest normalized CLIPScore (0.3076), confirming superior semantic alignment with text descriptions.

\begin{figure*}[!t]
    \centering
    \includegraphics[width=\linewidth]{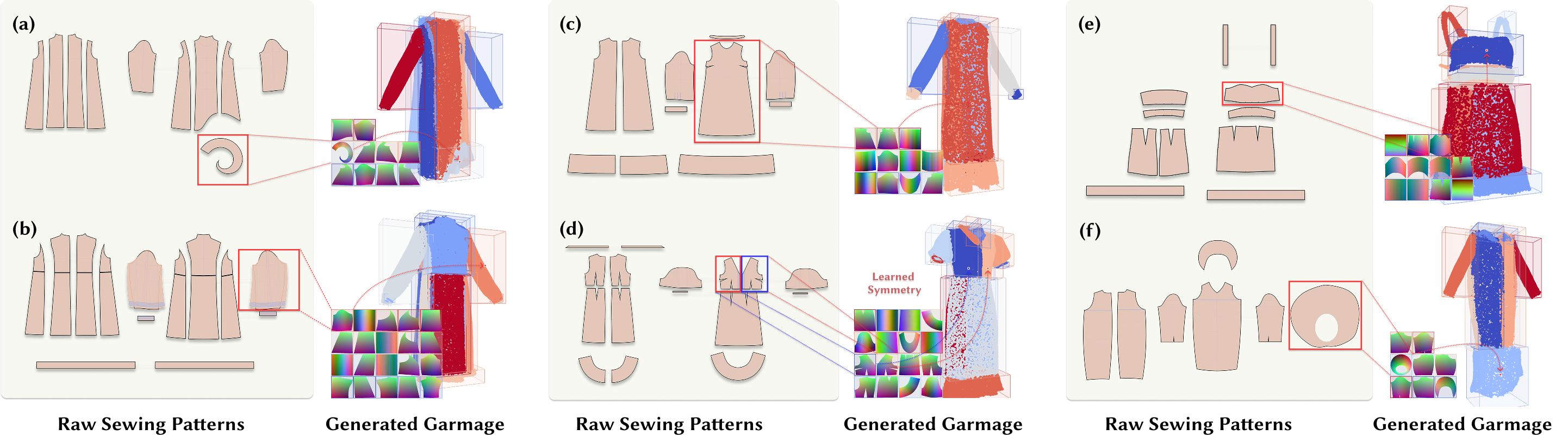}
    \caption{Automatic garment modeling from raw sewing patterns. Given flat sewing patterns without sewing relationship,  For clarity, we highlight several panels in the raw sewing pattern and label their corresponding generated Garmages with 3D point visualization.}
    \label{fig:cond_gen_sp}
\end{figure*}

\subsection{Automatic Garment Modeling}

Beyond its central role in gaming, virtual reality, and digital fashion, automatic garment modeling is equally critical for apparel manufacturing. By enabling manufacturers to visualize and validate sewing patterns before physical production, it helps reduce sampling costs, accelerate iteration, and minimize material waste. However, 3D garment modeling is traditionally a highly skill-demanding process, and the majority of industrial or internet-available resources contain only raw 2D sewing patterns without corresponding 3D assets. This gap underscores the necessity of an automatic pipeline capable of converting 2D patterns into faithful, simulation-ready 3D garments.
GarmageNet directly addresses this challenge by leveraging the masked latent encoding (Section~\ref{sec:latent-enc}) to provide fine-grained 3D initialization via the Garmage representation and establishes vertex-level stitching through GarmageJigsaw.

Figure~\ref{fig:cond_gen_sp} demonstrates garment generation results from raw sewing patterns. Several original panels are shown alongside their generated Garmages, highlighting that GarmageNet robustly handles automatic garment modeling even with unconventional designs such as hem panels (Figure~\ref{fig:cond_gen_sp} (a,f)). Moreover, although no explicit symmetry constraints were imposed during training, GarmageNet successfully captures symmetry cues inherent in the data (Figure~\ref{fig:cond_gen_sp} (d)), producing garments with strong bilateral consistency, particularly across sleeve panels.

\begin{figure*}[!t]
    \centering
    \includegraphics[width=\linewidth]{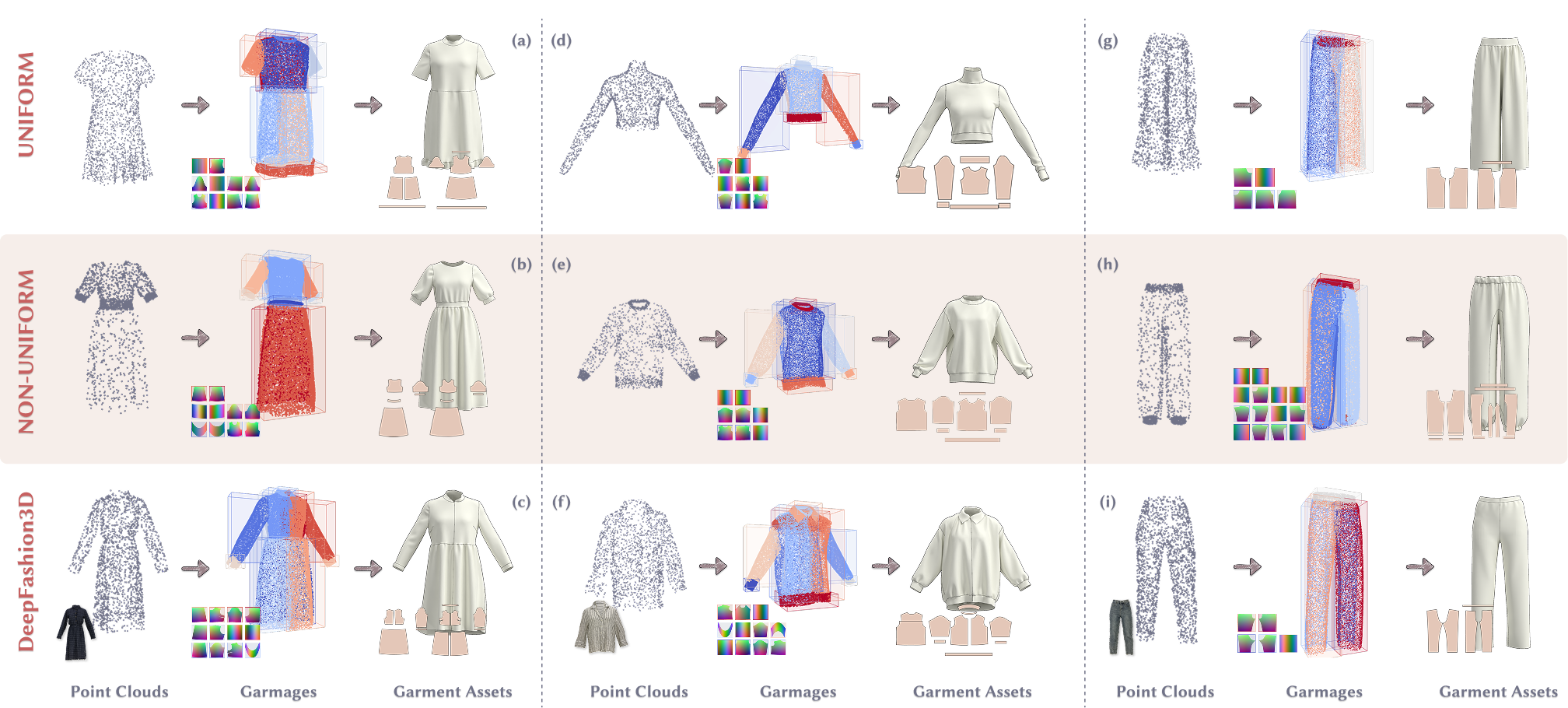}
    \caption{Point‐cloud‐conditioned garment synthesis with \textbf{GarmageNet}. Each row (a–f) shows: (left) an unstructured, sparse point cloud captured from a draped garment; (center) the generated \textbf{Garmage} representation—consisting of per‐panel geometry images (colored) and inferred panel contours (outlined); and (right) the final simulation‐ready 3D garment asset, obtained by vectorizing the extracted sewing patterns, recovering vertex‐wise stitches, and applying physics‐based draping. These results demonstrate GarmageNet’s ability to transform noisy, incomplete point clouds into fully structured sewing patterns and high‐fidelity draped garments. We note, however, that the network may be leveraging the non‐uniform sampling density of the input point clouds—implicitly revealing panel structure—to achieve these reconstructions.}
    \label{fig:cond_gen_pc}
\end{figure*}

\subsection{Sewing Pattern Recovery}

Advancements in 3D scanning and multi-view reconstruction technologies have greatly facilitated capturing realistic garment shapes, typically represented as unstructured point clouds. However, such raw 3D data lacks the structured information essential for garment production, thus necessitating effective methods for recovering structured sewing patterns from unstructured 3D representations.

GarmageNet addresses this critical industry challenge by accurately transforming point-cloud data of draped garments into structured Garmages, successfully recovering detailed sewing patterns. Figure ~\ref{fig:cond_gen_pc} showcases recovered sewing patterns, highlighting intricate folds and precise seam alignments. These recovered patterns closely match their original counterparts, demonstrating high accuracy in panel shapes, seam definitions, and adherence to industry production standards.

Qualitative analysis indicates that GarmageNet robustly identifies precise panel boundaries, seam connections, and garment folds from noisy input data, achieving reliable sewing pattern recovery even in complex garment configurations. This functionality positions GarmageNet uniquely within digital garment pipelines, effectively linking unstructured scan data to structured, production-ready garment assets, thereby significantly enhancing practical applicability in apparel manufacturing workflows.

\begin{figure}[!t]
    \centering
    \includegraphics[width=\linewidth]{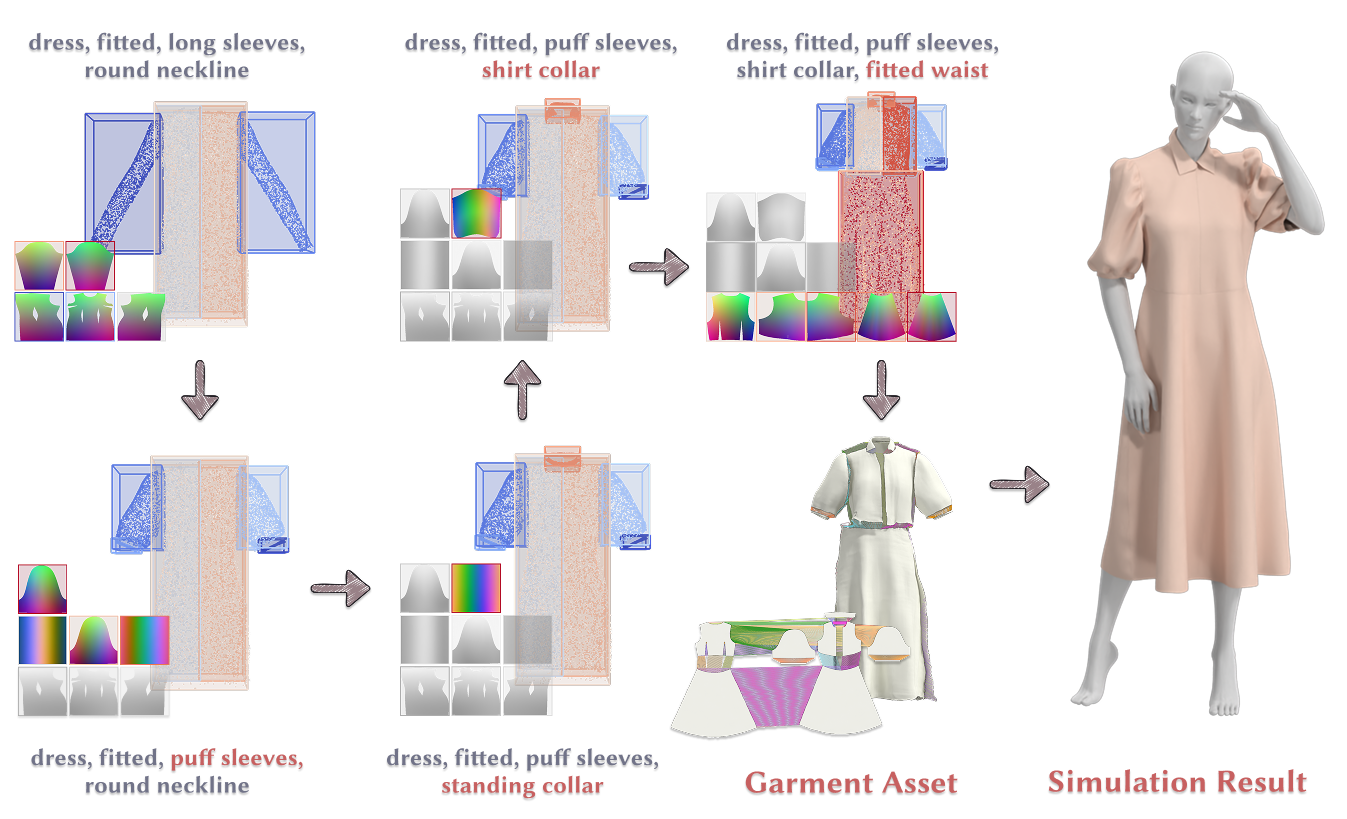}
    \caption{Interactive garment editing using conventional design instructions. Starting from an initial Garmage (top-left), users issue sequential edits—e.g.\ replacing a round neckline with a shirt collar, adding a fitted waist, and switching to a standing collar—while all unchanged panels remain in grey and only the edited panels (in color) are updated in their geometry images. Each intermediate Garmage is decoded into a full garment asset and re-simulated, demonstrating how our framework seamlessly incorporates standard pattern-making edits into the generation and draping pipeline.}
    \label{fig:cond_editing}
\end{figure}

\subsection{Progressive Generation and Editing}

Beyond generating garments directly from text prompts, our framework also supports advanced garment editing functionalities, such as adding, deleting, or replacing components of an existing garment. This capability significantly enhances the flexibility of the design process, allowing designers to iteratively refine garments based on new inputs while preserving key structural features.

Recall from Eq.~\ref{eq:garmage_repr} that a Garmage consists of a set of panels, each represented by a 2D dimension $D_i$, a 3D axis-aligned bounding box $B_i$, and a normalized geometry image patch $I_i$. After generating a garment using text prompts, users can modify the original prompts to reflect desired changes, such as removing or replacing specific garment components.

When a text prompt is updated, the garment is regenerated, and the newly generated panels are compared against the original panels based on their 2D dimensions and 3D bounding boxes. Panels with high similarity are marked for retention, while those with low similarity are flagged for modification. 
The editing process is akin to inpainting in image generation models, where only the panels requiring modification are regenerated. Retained panels are treated in a way similar to diffusion-based denoising, where the original features are preserved and augmented with noise according to the current timestep, guiding the model to retain the established characteristics of those panels. In this way, the modifications are localized to the relevant areas of the garment without disrupting the overall design, and new panels are generated in a manner that ensures smooth transitions at the interfaces between modified and retained panels (e.g., sleeve holes), maintaining coherence in both geometry and design.

Figure~\ref{fig:cond_editing} illustrates an progressive generation process where we first generate a \textit{fitted, sleeveless dress} from text prompts (a), then add long sleeves to the dress (b), and modify the sleeves to puff sleeves (c). Next, we add standing lapel collars to the dress (d) and modify them to shirt collars (e). Finally, if the user is dissatisfied with the generated result, we can even modify the entire initial bodice panels (f). Note that the structural integrity and stylistic consistency are preserved during the whole editing process.

\section{Conclusion}
\label{sec:conclusion}
In this work, we introduced \emph{GarmageNet}, the first end-to-end framework for unified 2D–3D garment synthesis. At its core is \emph{Garmage}, a novel panel-aligned geometry-image representation that encodes both discrete sewing-pattern structure and continuous draping geometry into a compact, image-based format. By training a latent-diffusion transformer on Garmage tokens, our approach enables both unconditional and conditional generation from diverse design modalities—including text, sketches, point clouds, and raw sewing patterns—while preserving fine-grained panel topology and delivering high-fidelity, simulation-ready initializations. We further presented \emph{GarmageJigsaw}, a dedicated module that leverages 2D silhouettes and 3D spatial cues to recover vertex-level sewing relationships, seamlessly converting generated Garmages into vectorized sewing patterns and triangulated meshes compatible with physics-based simulation.

To support this framework, we introduced \emph{GarmageSet}, the largest garment dataset to date built from real production data. It contains 14,801 professionally drafted sewing patterns paired with fine-grained 3D assets, detailed structural and style annotations, and multi-view renderings with point clouds. By bridging industrial-grade data quality with rich multi-modal supervision, GarmageSet provides a robust foundation for advancing garment modeling research toward deployable manufacturing pipelines.

Comprehensive evaluations demonstrate that GarmageNet outperforms state-of-the-art structure-centric and geometry-centric approaches in terms of quality, diversity, and robustness, and achieves higher simulation success rates compared to rigid and optimization-based automatic initialization methods.

\begin{figure}[!t]
    \centering
    \includegraphics[width=\linewidth]{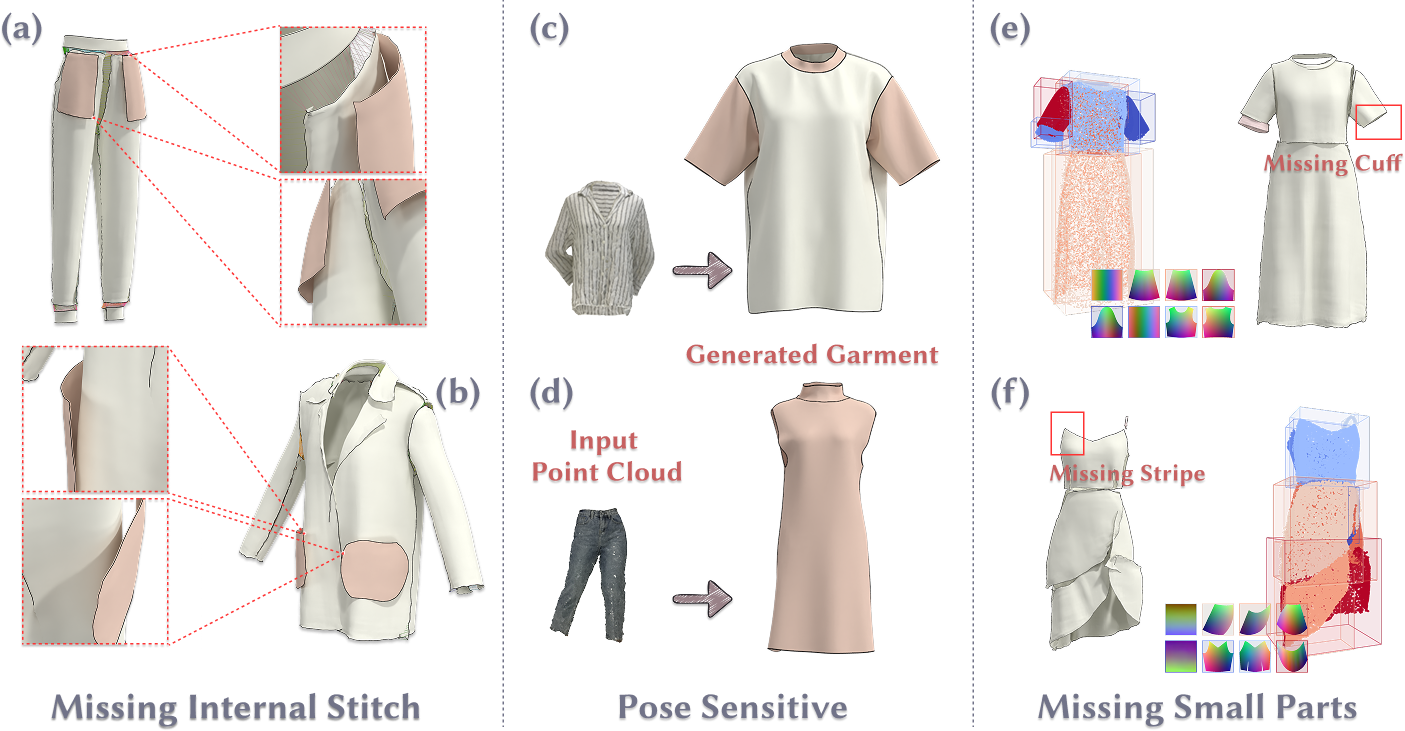}
    \caption{Failure cases of our proposed framework. (a–b) Missing internal seams: while interior stitch lines and pocket attachments are present in training, GarmageJigsaw receives only boundary points at inference, preventing correct stitching of pocket-like panels. (c–d) Pose sensitivity: since GarmageSet contains only garments draped on a size-S A-pose avatar, point-cloud–conditioned generation can fail under pose variations. (e–f) Missing small or auxiliary parts: our zero-padding scheme may cause the loss of small panels such as cuffs or thin stripes.}
    \label{fig:limitations}
\end{figure}

\section{Limitations and Future Work}
\label{sec:limitation}
While GarmageNet demonstrates strong performance in unified garment modeling, several limitations remain:

\paragraph{Lacking body shape and pose variance} To control dataset preparation costs, GarmageSet currently contains garments drafted only for a standard size-S avatar posed in A-pose. As a result, GarmageNet may be sensitive to pose or shape variations, particularly under point-cloud–conditioned generation (Figure~\ref{fig:limitations} (c,d)). Incorporating body-shape and pose diversity is a priority for future releases.

\paragraph{Handling of interior seams and pocket-like components} Although GarmageNet and GarmageJigsaw are trained with interior seams and panels that attach along them (for instance, pocket panels in Figure~\ref{fig:limitations} (a,b)). During inference, we only pass boundary facets to GarmageJigsaw to identify sewing relationships between those boundary facets. This limitation prevents the model from stitching  the generated pocket panels to bodice panels due to the absence of interior edges on bodice panels (Figure~\ref{fig:limitations} (b)).

\paragraph{Multi-layer designs and small-panel handling} Because GarmageSet includes garments with layered collars and sleeve designs, GarmageNet occasionally generates redundant sleeve panels, leading to reduced accuracy in panel-count metrics (\#Panels) as shown in Table~\ref{tab:cross_dataset_comp}. Moreover, the zero-padding strategy adopted during training can cause small or thin panels (e.g., cuffs or narrow stripes) to be omitted (Figure~\ref{fig:limitations} (c)).

\paragraph{Future Work} We plan to expand \textbf{GarmageSet} to include a broader spectrum of body shapes and poses beyond the current size-S A-pose, thereby reducing sensitivity to pose and shape variations in conditional generation. We also intend to enhance seam representations by adding interior seam annotations and pocket-attachment edges, enabling \textbf{GarmageJigsaw} to accurately handle pocket-like components. Furthermore, we will continue to refine the \textbf{Garmage} representation and optimize the architectures of GarmageNet and GarmageJigsaw to better accommodate multi-layered designs and small panels.

\begin{acks}
This work was supported by Key R\&D Program of Zhejiang (No. 2023C01047). We also gratefully acknowledge Shanmei Shen, Jing Chao, Fuqi Wang and the Style3D Digital Product Creation (DPC) team for their contributions in building GarmageSet.
\end{acks}

\bibliographystyle{ACM-Reference-Format}
\bibliography{sample-base}

\end{document}